\documentclass[journal]{IEEEtran}
\usepackage{upgreek}
\usepackage{graphicx}
\usepackage{subfigure}
\usepackage{amsmath,bm}
\usepackage{amssymb}

\usepackage{pifont}
\usepackage{enumerate}

\usepackage{url}

\usepackage[square, comma, sort&compress, numbers]{natbib}

\usepackage{color}

\begin{document}

\title{  Performance of a High Power and Capacity\\ Mobile SLIPT Scheme}
\author{
	Mingliang Xiong~\IEEEmembership{Graduate  Student Member,~IEEE}, Qingwen Liu,~\IEEEmembership{Senior Member,~IEEE}, \\Shengli Zhou,~\IEEEmembership{Fellow,~IEEE}, Shun Han, and Mingqing Liu

\thanks{
The corresponding author: Qingwen Liu.
}
\thanks{This work was supported in part by the National Natural Science Foundation of China under Grant  62071334, in part by the National Key Research and Development	Project under Grant 2020YFB2103900 and Grant 2020YFB2103902, in part by the Shanghai Municipal Science and Technology Major Project under Grant 2021SHZDZX0100, in part by the Shanghai Municipal Commission of Science and Technology Project under Grant 19511132101, in part by the Natural Science Foundation of Shanghai under Grant 22ZR1462900, and in part by the Fundamental Research Funds for the Central Universities 22120210543.}
\thanks{
	M. Xiong, Q. Liu, S. Han, and M. Liu
	are with the College of Electronics and Information Engineering, Tongji University, Shanghai 201804, China (e-mail: xiongml@tongji.edu.cn;    qliu@tongji.edu.cn; nuisthanshun@outlook.com;  clare@tongji.edu.cn). S. Zhou is with Department of Electrical and Computer Engineering, University of Connecticut, Storrs, CT 06250, USA (e-mail: shengli.zhou@uconn.edu).}
		
}

\maketitle

\begin{abstract}
The increasing demands of power supply and data rate for mobile devices promote the research of simultaneous wireless information and power transfer~(SWIPT). Optical SWIPT, as known as simultaneous light information and power transfer~(SLIPT), has the potential for providing high-capacity communication and high-power wireless charging.  However, SLIPT technologies based on light-emitting diodes have low efficiency due to energy dissipation over the air. Laser-based SLIPT technologies need strict positioning accuracy and scanning resolution, which may lead to the increase of costs and complexity. In this paper, we propose a mobile SLIPT scheme based on spatially separated laser resonator~(SSLR) and intra-cavity second harmonic generation. The power and data are transferred via separated frequencies, while they share the same self-aligned resonant beam path, without the needs of receiver positioning and beam steering. We establish the analysis model of the resonant beam power and its second harmonic power.  Numerical results show that the proposed system can achieve watt-level battery charging power and above $10$-bit/s/Hz achievable rate at $6$-m distance, which satisfies the requirements of most indoor mobile devices.
\end{abstract}


\section{Introduction}\label{sec:intro}

\IEEEPARstart{T}{he} increasing demands of virtual reality~(VR), augmented reality~(AR), and other bandwidth-consuming services bring many challenges to conventional mobile networks, including radio frequency spectrum crisis and fast battery depletion.  Simultaneous wireless information and power transfer~(SWIPT) technology is an attractive way to cope with these challenges, as it can provide sustainable power supply through the communication link~\cite{Varshney2008}. In the sixth-generation (6G) mobile network, SWIPT service is highly expected~\cite{a180820.09}. The final form of mobile network can be completely different from what we are experiencing today~\cite{a210620.01}.
   
  \begin{figure}[t]
  	\centering
  	\includegraphics[width=3.2in]{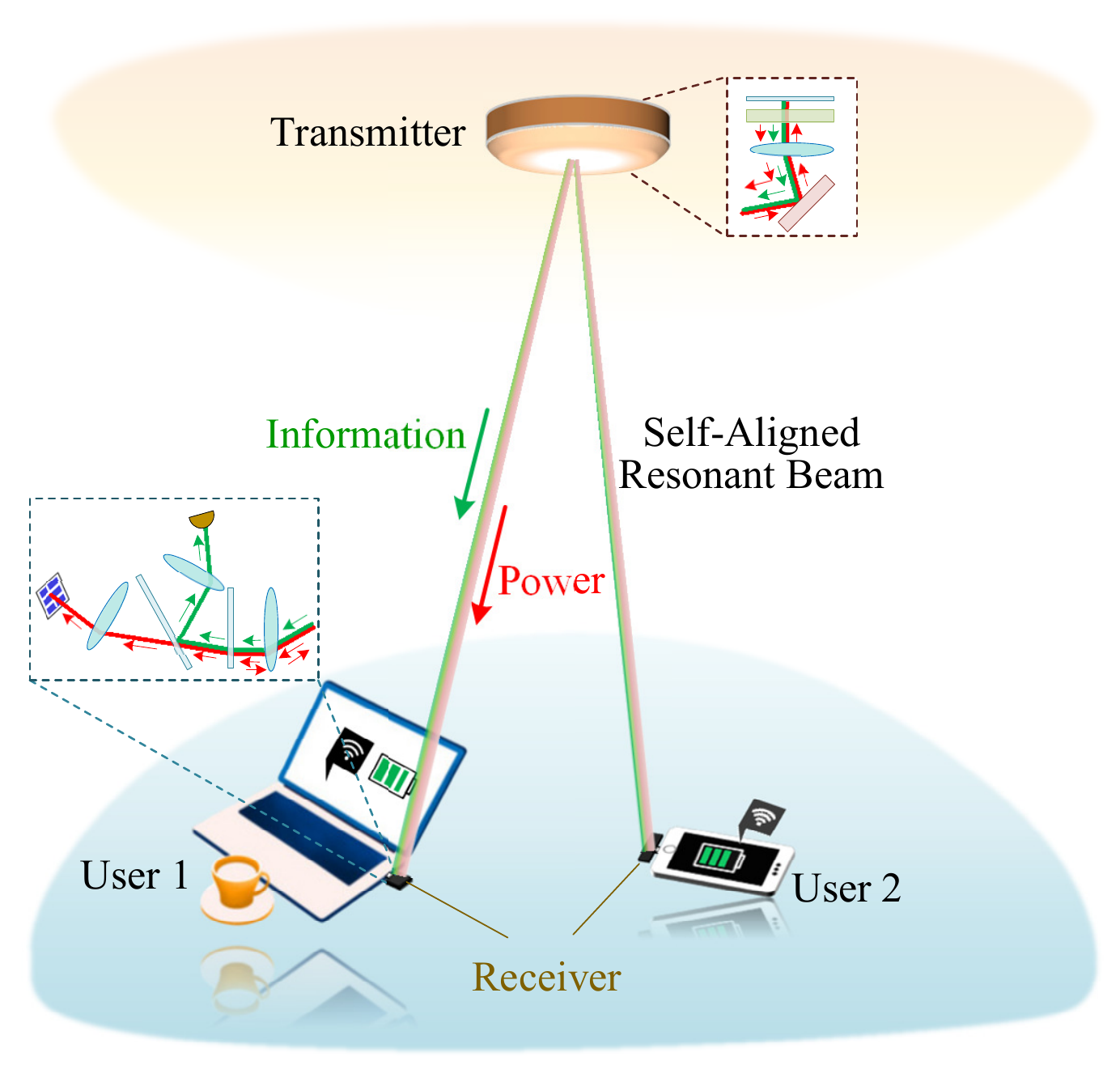}
  	\caption{Scenario of mobile optical SWIPT based on resonant beam}
  	\label{fig:appli}
  \end{figure}
  
SWIPT using radio-frequency~(RF) carrier has been well-studied. RF wave can provide a large coverage area, non-line-of-sight~(NLoS) transmission, low scattering and absorption in atmosphere, and mature manufacturing; hence, it has many application scenarios such as wireless sensor network distributed in a large area. However, the shortage of RF spectrum resources limits the development of the RF-based SWIPT. Owing to high communication bandwidth, simultaneous light information and power transfer~(SLIPT) has arisen much attention.
Freespace optical communications~(FSO) technologies have the potential to support over $1$~Tbit/s data rate~\cite{Ciaramella2009}, which points out the promising prospect of SLIPT. Paper~\cite{Ma2019} presented a novel SLIPT, in which a solar cell is employed for energy harvesting~(EH) and a photodiode~(PD) is used for information detection. The SLIPT system presented in~\cite{a210901.04} splits the direct-current (DC) energy and the alternating-current (AC) information signal from the received power by a separator; it is also optimized with new strategies and policies for the balance of the trade-off between EH and the quality of service~(QoS). The work in~\cite{a211109.03}
presented a promising relaying SLIPT system with rate optimization, in which the powers for EH and communication are split directly by a power splitter. Paper~\cite{sandalidis2017} investigated the use of VLC links for simultaneous lighting, communications, and energy harvesting. Paper~\cite{li2017sum} investigated the sum rate maximization problem in a downlink SLIPT system and proposed a low-complexity iterative algorithm.
 Performance comparison between RF-based SWIPT and SLIPT was carried out in~\cite{a211109.01}, which demonstrated the superior performance of SLIPT. The transmitters employed in these SLIPT systems are-light emitting diodes (LEDs).

However, transmitters with isotropic antennas or LEDs exhibit very low transmission efficiency due
to the path loss over the large divergence angle. For example, an isotropic antenna operating at $900$~MHz and supplied with $4$ W  can only transfer $5.5$~$\mu$W to the receiver at $15$-m distance~\cite{a210620.02}. From LED radiation within $120^\circ$ coverage angle, the receiver can only obtain $1.4$-mW electric power at $1.5$-m distance~\cite{a210620.03}. Therefore, for power-hungry devices, narrow-beam carrier is preferred. For instance, Iyer \textit{et~al.} demonstrated a safeguard-beam-based laser power transfer system which can  safely deliver over $2$-W electrical power at a range of $4.3$~m~\cite{a180805.04}. Rizzo \textit{et~al.} employed $10$-W $976$-nm laser as the transmitter, achieving $2.37$-W electrical output at the receiver~\cite{a210621.01}. Furthermore, hybrid optical wireless power and data
transmission system is also introduced in~\cite{a210621.02}, which employed two laser diodes at different frequencies to transfer power and information independently;  this experimental work achieved $150\mbox{-mW}$  received optical power and $10$-MHz bandwidth. These laser-based optical SWIPT technologies have to obtain the position of the receiver and then direct the laser beam to the receiver. Although beam steering can be realized by many non-mechanical devices, such as two-dimensional (2D) fiber arrays~\cite{a191111.01}, in-fiber diffraction gratings~\cite{a201130.03}, optical phased
arrays~(OPAs)~\cite{a201201.02}, crossed gratings~\cite{a190611.03}, and spatial light modulators~(SLMs)~\cite{a190514.01}, the strict needs of positioning accuracy and scanning resolution may lead to the increase of costs and complexity, especially for long-range transmission~\cite{a211115.01}. Particularly, for watt-level laser transfer, inaccurate positioning and insensitive detection will lead to dangerous events.

\begin{figure*}
	\centering
	\includegraphics[width=5.6in]{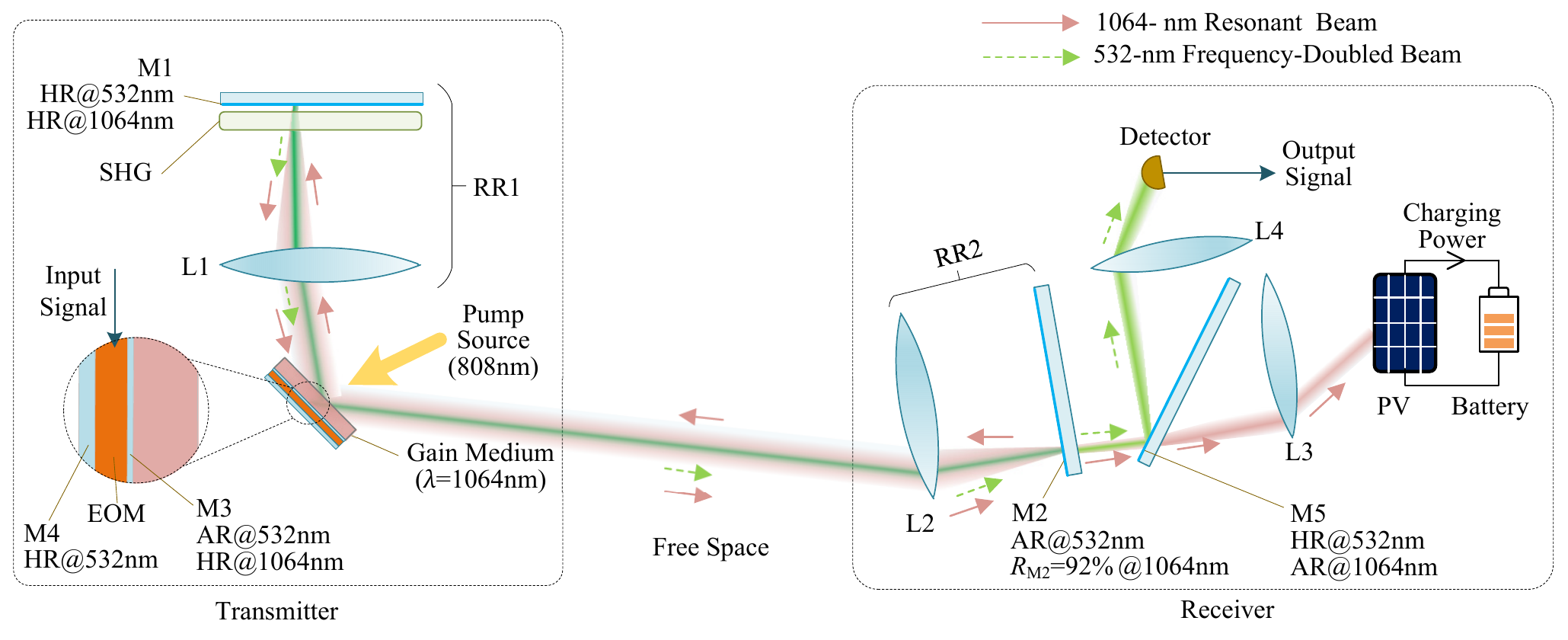}
	\caption{Diagram of system structure (M1 and M2: mirrors; M3 and M4: coatings; M5: dichroic mirror; L1 to L4: lenses; RR1: the retroreflector at the transmitter, consisting of M1 and L1; RR2: the retroreflector at the receiver, consisting of M2 and L2; SHG: second harmonic generator; EOM: electro-optical modulator; PV: photovoltaic panel; AR: anti-reflective coating, i.e., transmittance $\geqslant 99.5$\%; HR: high-reflective coating, i.e., reflectivity $\geqslant 99.5$\%; M2 is partially reflective at $1064$~nm with reflectively of $R_{\rm M2}$)}
	\label{fig:design}
\end{figure*}

Resonant beam generated by spatially separated laser resonator~(SSLR) can be used for mobile optical SWIPT without the requirements of positioning and beam steering, as shown in Fig.~\ref{fig:appli}. Recently, high-power mobile wireless charging using resonant beam was proposed in~\cite{a180727.01}, which is known as resonant beam charging~(RBC) or distributed laser charging~(DLC). Experiment on RBC was demonstrated in~\cite{RBCexperiment}, which received $0.6$-W electrical power at $2$-m distance. Resonant beam communication~(RBCom) system with high-capacity channel was illustrated in~\cite{a191111.06}. The primary structure of SSLR origins from Lindford's very long laser cavity which has two oppositely-placed corner cube retroreflectors~\cite{a190318.02}. The safety  of the SSLR was verified in~\cite{WFang2021}. The SSLR also has limitations with respect to communications, as the signal inside the cavity bounces back and forth, leading to echo interference. To address the echo interference issue, SSLR structures with external-cavity second harmonic generation (SHG) and intra-cavity SHG are proposed in~\cite{MXiong2021} and \cite{MXiong2021.2}, respectively. However, \cite{MXiong2021} and \cite{MXiong2021.2} only consider information transfer, ignoring the ability of the resonant beam in power transfer. \cite{a210823.01} and \cite{MLiu2021} investigated a SWIPT system based on corner-cube-based SSLR and cat's-eye-based SSLR, respectively, while they transfer information and power via the same wavelength, and split the power proportionally at the receiver. In this case, the modulation applied to the beam reduces the charging power, and hence, the power loss in this process is inevitable. Moreover, they did not address the echo interference issue.


In this paper, we propose a resonant beam SWIPT~(RB-SWIPT) system based on SSLR and intra-cavity SHG. In addition to the aforementioned advantages of the RBC and RBCom system in power, mobility, and capacity, the RB-SWIPT system separates the communication beam from the power transfer beam (i.e., the resonant beam) at the transmitter and converts them into  signal and electricity at the receiver independently using a PD and a photovoltaic panel (PV).

The contributions of our work are as follows.
\begin{enumerate}
	\item[\bf 1)] We propose the SHG-based RB-SWIPT system to provide a high-power and high-capacity mobile transmission channel, and design an structure for optical frequency-splitting transmission and independent conversion. In this design, the modulation is only applied to the information transfer beam, and thus, it cannot affect the power transfer beam, which avoids  extra power loss.
	\item[\bf 2)] We establish the coupling model of the resonant beam power and its second harmonic power and analyze the performance of the RB-SWIPT system on the received charging power and achievable rate.
\end{enumerate}

The remainder of this paper is organized as follows. Section~\ref{sec:design} presents the system design of the RB-SWIPT system. Section~\ref{sec:model} illustrates the system analysis model. Section~\ref{sec:result} demonstrates the results of performance evaluation. Section~\ref{sec:discu} discusses the concerns on  safety,  energy efficiency, and  misalignment. Finally, we conclude in Section~\ref{sec:con}.

\section{The Proposed System}
\label{sec:design}

The SWIPT system is based on the SSLR and SHG scheme presented in~\cite{MXiong2021.2}. As  demonstrated in Fig.~\ref{fig:design}, power is transferred via  $1064$-nm resonant beam, while information is transferred via  $532$-nm frequency-doubled beam. The frequency-doubled beam is generated from the resonant beam by an SHG crystal and modulated by an electro-optic modulator~(EOM). At the receiver, part of the resonant beam is allowed to pass through mirror M2, and then is focused on the PV by lens L3. At last, the extracted resonant beam is converted into current  by the PV to charge the battery. The frequency-doubled beam passes through M2, and then is separated from the resonant beam by a dichroic mirror M5; and finally, it is focused on the PD by lens L4 and converted into current signal by the PD.

\begin{figure}
	\centering
	\includegraphics[width=2.2in]{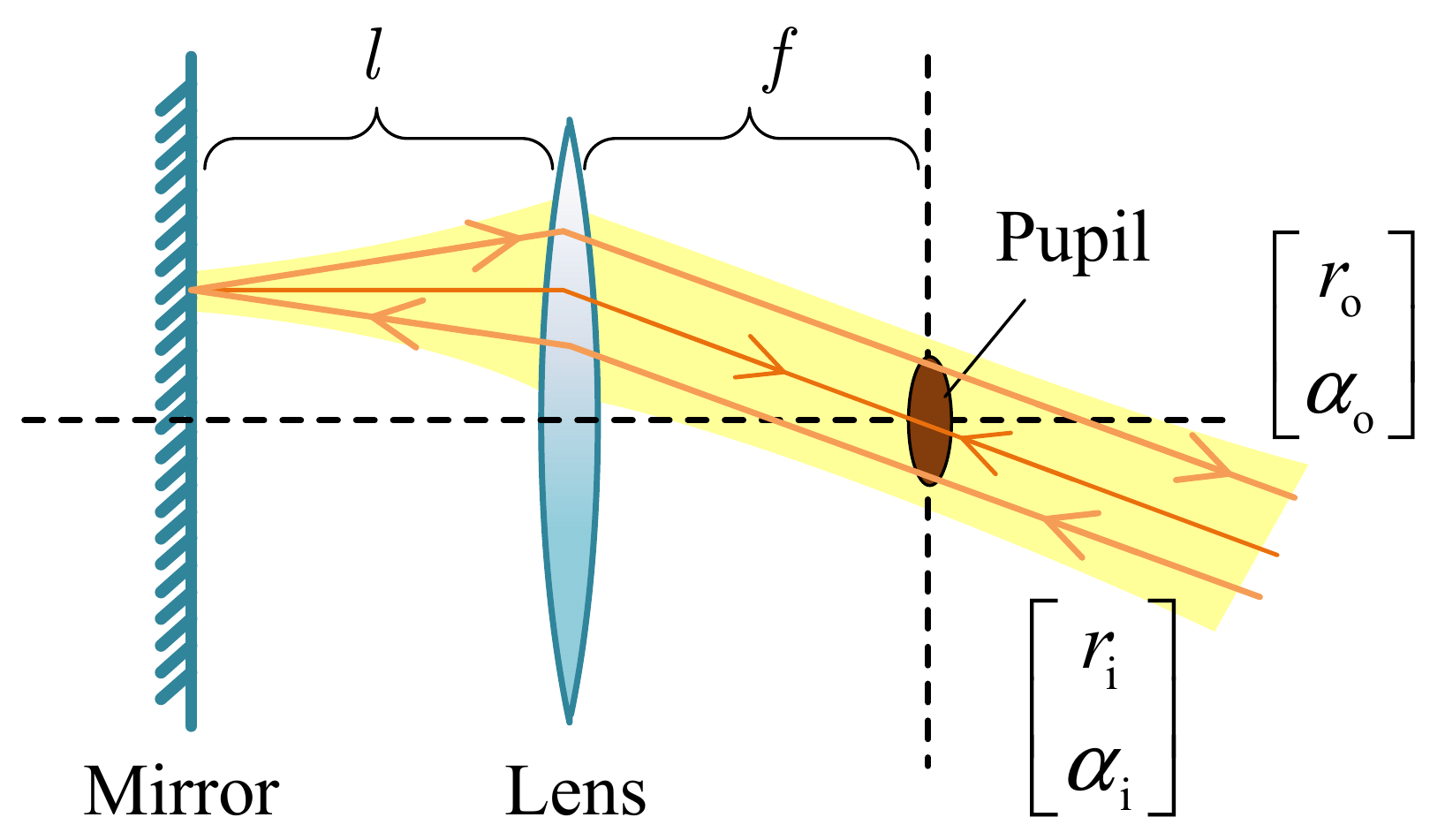}
	\caption{Telecentric cat's eye retroreflector}
	\label{fig:retro}
\end{figure}

The resonant beam is generated inside the SSLR cavity. The SSLR consists of two TCRs RR1 and RR2 which are placed at the transmitter and the receiver, respectively. Retroreflectors are devices to reflect the incident beam back to its incoming direction. Therefore, photons inside the SSLR cavity can oscillate between RR1 and RR2 regardless of the retroreflectors' location, as long as these TCRs are in each other's field of view~(FOV). Generally, a TCR consists of a lens and a rear mirror which are formed coaxially and separated with a space.
The ray-transfer matrix of the telecentric cat's eye retroreflector is expressed as \cite{MXiong2021.2}
\begin{align}
\mathbf{M}_{\rm RR}&=			\begin{bmatrix}
1&f\\0& 1
\end{bmatrix}
\begin{bmatrix}
1&0\\-1/f& 1
\end{bmatrix}
\begin{bmatrix}
1&l\\0& 1
\end{bmatrix}
\begin{bmatrix}
1&0\\0& 1
\end{bmatrix}
\begin{bmatrix}
1&l\\0& 1
\end{bmatrix}
\nonumber
\\
&~~~~
\begin{bmatrix}
1&0\\-1/f& 1
\end{bmatrix}
\begin{bmatrix}
1&f\\0& 1
\end{bmatrix}
\nonumber
\\
&=
\begin{bmatrix}
1&0\\-1/f_{\rm RR}& 1
\end{bmatrix}
\begin{bmatrix}
-1&0\\0 &-1
\end{bmatrix},
\end{align}
and
\begin{equation}
f_{\rm RR}=\dfrac{f^2}{2(l-f)},
\label{equ:fRR}
\end{equation}
where $f$ is the focal length of the lens, and $l$ is the interval between the lens and the mirror. If $l=f$, $\mathbf{M}_{\rm RR}$ represents an ideal TCR. The ray-transfer matrix of an optical system is used to calculate the position and the transfer direction of the output ray from those of the input ray. Given the vector,  $[r_{\rm i}, \alpha_{\rm i}]^{T}$, of the input ray  of the ideal TCR, we can obtain the vector of the output ray; that is
\begin{equation}
\begin{bmatrix}
r_{\rm o}\\
\alpha_{\rm o}
\end{bmatrix}
=\mathbf{M}_{\rm RR}\bigg|_{l=f}
\begin{bmatrix}
r_{\rm i}\\
\alpha_{\rm i}
\end{bmatrix} =\begin{bmatrix}
-r_{\rm i}\\-\alpha_{\rm i}
\end{bmatrix},
\label{equ:rayout}
\end{equation}
where $r_{\rm i}$ ($r_{\rm o}$) and $\alpha_{\rm i}$ ($\alpha_{\rm o}$) are the transverse displacement and the angle of the input (output) ray, respectively, with respect to the optical axis. From~\eqref{equ:rayout}, we can see that the output ray is parallel to the input ray, as shown in Fig.~\ref{fig:retro}. Nevertheless, in SSLR, $l$ should be a little greater than $f$ to ensure a stable resonator, which is illustrated in the following section. TCR has a pupil located at the focal point of the lens. Only the beams that pass through the pupil of the TCR can be retro-reflected. Therefore, the intra-cavity resonant beam always passes through the pupils of RR1 and RR2. We place the gain medium at the pupil of RR1, so that the resonant beam can always reach the gain medium regardless of the location of the receiver, as long as the receiver is in the FOV of the transmitter, and vice versa. Similarly, the EOM is placed behind the gain medium, so that the frequency-doubled beam can always reach the EOM.

In this paper, we employ an neodymium-doped yttrium orthovanadate (Nd:YVO$_4$) crystal as the gain medium. This material can amplify $1064$-nm light under the irradiation of $808$-nm pump source light. It can also be pumped with $880$-nm light, which helps reduce the heat dissipation, but a thicker crystal is needed to compensate for the decreased absorption coefficient. The choice of the pump wavelength depends on the FOV requirement and the heat-sinking capability. As the gain medium absorbs the pump power, the atoms at the lower energy level transit to the the upper level continuously, providing the stimulated emission capability. The intra-cavity resonant beam oscillating between RR1 and RR2 is amplified by the gain medium through the stimulated emission. In the stimulated emission transition of an excited atom at the upper level, an extra photon identical to the input one is generated, and meanwhile, the excited atom transits to the lower level. Since the loss of the resonant beam is compensated by the gain, the resonance in the cavity can be maintained. Much heat is generated by the gain medium, so it is important to  design the heat sink well. For example, we can use indium piece to fill the gap between the gain medium (with the EOM) and the heat sink. For high-power pump, a cooling module is helpful for the heat dissipation. SHG crystals are devices to generate frequency-doubled beam from a fundamental beam. In this system, the fundamental beam for SHG is the resonant beam. We place the SHG crystal inside RR1 and make it parallel to the rear mirror. Because beams that enter through the pupil of the TCR are perpendicular to the rear mirror, they have the same incident angle to the SHG crystal. This feature meets the phase-matching requirement of the SHG crystal.

Various anti-reflective~(AR), high-reflective~(HR), and partial-reflective coatings at different wavelengths are employed. M1 is coated with HR coating at $1064$~nm. M2 is coated with partial-reflective coating at $1064$~nm. Hence, M1 and M2 capture most of the resonant beam power. Only a small part of the resonant beam power is released by M2 for battery charging. The back side of the gain medium is coated with HR coating at $1064$~nm to prevent the resonant beam passing through and being changed by the EOM; and it is also coated with AR coating at $532$~nm to allow the frequency-doubled beam going into the EOM. M1 and the back side of the EOM are coated with HR coatings at $532$~nm to reflect the frequency-doubled beam. M2 is coated with AR coating at $532$~nm to allow the frequency-doubled beam going out of the cavity, preventing any possible intra-cavity oscillation at $532$~nm. The front side of M5 is coated with  HR coating at $532$~nm and AR coating at $1064$~nm, so that M5 can separate the two wavelengths. Besides, the two sides of each lens and the SHG crystal, the input plane of the detector and the PV, and the back side of M5 are coated with AR coatings at both $1064$~nm and $532$~nm to  prevent unnecessary reflection.

\section{System Analysis}
\label{sec:model}

The proposed SWIPT system has two branches, i.e., the power transfer~(PT) branch and the information transfer~(IT) branch. In this section, we at first introduce the  mechanism of resonant beam generation, and present the necessary condition for a stable SSLR. Then, the process of communication carrier generation is illustrated. At last, the photoelectric conversion and battery charging as well as the detection of the information signal are described.

\subsection{Resonant Beam Generation}
\begin{figure}[t]
\centering
\includegraphics[width=3.4in]{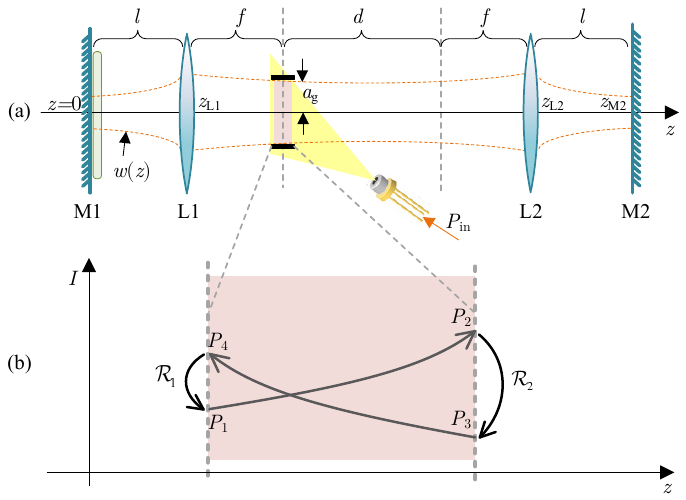}
\caption{System model for (a) intra-cavity beam radius and (b) intra-cavity circulating power}
\label{fig:model}
\end{figure}

In our system, the power is transferred via the resonant beam whose frequency is $\nu$. As shown in Fig.~\ref{fig:model}(a), the resonant beam is generated between M1 and M2. In this work, we only consider that RR1 and RR2 have exactly the same  parameters, including the same focal length $f$ and the same space interval $l$.  The beam radius $w(z)$ varies along the $z$-axis. The SSLR can be expressed by a single-pass ray-transfer matrix, which is also called the ABCD matrix; that is~\cite{MXiong2021}
\begin{align}
\begin{bmatrix}
A&B\\C&D
\end{bmatrix}
=&	\begin{bmatrix}
1&0\\0&1
\end{bmatrix}
\begin{bmatrix}
1&l\\0&1
\end{bmatrix}
\begin{bmatrix}
1&0\\ -1/f&1
\end{bmatrix}
\begin{bmatrix}
1&2f+d\\0&1
\end{bmatrix}
\nonumber
\\
&\begin{bmatrix}
1&0\\ -1/f&1
\end{bmatrix}
\begin{bmatrix}
1&l\\0&1
\end{bmatrix}
\begin{bmatrix}
1&0\\0&1
\end{bmatrix}\nonumber\\
=&
\begin{bmatrix}
-1-\dfrac{d}{f}+\dfrac{dl}{f^2} &
2f-2l+d-\dfrac{2dl}{f}+\dfrac{dl^2}{f^2}\\
\dfrac{d}{f^2}&
-1-\dfrac{d}{f}+\dfrac{dl}{f^2}
\end{bmatrix},
\end{align}
where $d$ is the transmission distance that is defined as the space interval between the pupils of RR1 and RR2. The elements of ABCD matrix can be used to analyze the stability of resonators~\cite{a200224.01,a190511.01}. Let $g_1^*=A$ and $g_2^*=D$, we can use the condition $0<g_1^*g_2^*<1$ to judge the stability of the SSLR. Here, using $f_{\rm RR}$ expressed in~\eqref{equ:fRR}, we can know that the stable-cavity condition is $0\leqslant d \leqslant 4f_{\rm RR}$. Note that the real distance also depends on the losses in the cavity. Moreover, high temperature at the gain medium will result in thermal lens, which will reduce the range of stable regime. Therefore, a cooling system design for the gain medium is important.

The components of the ABCD matrix can also be used to estimate the beam radius at arbitrary location along the optical axis. We should at first obtain the $q$-parameter and then use it to calculate the beam radius. There are multiple transverse modes included in the resonant beam; and their superposition determines the radius of the resonant beam. The modes that can be included in a resonant beam are determined by the apertures of the optical devices in the cavity. Among these modes, the fundamental mode TEM$_{00}$ exhibits the smallest radius. We can obtain the radius of the fundamental mode at location $z$ by~\cite{a181221.01}
\begin{equation}
w_{00}(z)=\sqrt{-\dfrac{\lambda}{\pi\Im\left[1/q(z)\right]}},
\end{equation}
where $\Im[\cdot]$ takes the imaginary part of a complex number, and $\lambda$ is the wavelength of the resonant beam. The $q(z)$ parameter is computed by~\cite{MXiong2021}
\begin{equation}
q(z)=
\left\{
\begin{array}{ll}
j |L^*|\sqrt{\dfrac{g_2^*}{g_1^*(1-g_1^*g_2^*)}}+z,&z\in[0,z_{\rm L1}] \vspace{2ex}\\
\dfrac{q(z_{\rm L1})}{{-q(z_{\rm L1})}/{f}+1}+(z-z_{\rm L1}),&z\in(z_{\rm L1},z_{\rm L2}] \vspace{2ex}\\
\dfrac{q(z_{\rm L2})}{{-q(z_{\rm L2})}/{f}+1}+(z-z_{\rm L2}),&z\in(z_{\rm L2},z_{\rm L3}] \vspace{2ex}\\
\dfrac{q(z_{\rm L3})}{{-q(z_{\rm L3})}/{f}+1}+(z-z_{\rm L3}),&z\in(z_{\rm L3},z_{\rm PV}] \\
\end{array}
\right.
\label{equ:qParam}
\end{equation}
where $L^*=B$, $z_{\rm L1}=f$, $z_{\rm L2}=l+2f+d$, $z_{\rm L3}=3l+2f+d$, and $z_{\rm PV}=3l+3f+d$ are the locations of lens L1, lens L2, lens L3, and PV, respectively. Note that L3 and L4 are symmetric about M5. Also, the positions of the PD and the PV are symmetric about M5. From~\eqref{equ:qParam}, we can observe that $q(z)$ is divided into three parts along $z$-axis. All the $q$-parameters are derived from $q(0)$. In fact, the $q$-parameters at one side of an optical element is derived from the $q$-parameters at the other side using the transfer-matrix method.

The ratio of the beam radius to the TEM$_{00}$ mode radius is a constant along the $z$-axis. This ratio is called the beam propagation factor. To compute the beam radius $w(z)$ at arbitrary location $z$, we can firstly obtain the resonant beam radius $w(l+f)$ and the TEM$_{00}$ mode radius $w_{00}(l+f)$ at the gain medium  and then calculate the beam propagation factor. Since the aperture of the gain medium has the smallest aperture among the devices inside the cavity, we can consider that high-order modes whose radius is greater than the gain medium aperture are blocked, while only those with smaller radius remain in the cavity. Therefore, we can make a good approximation that $w(l+f)\approx a_{\rm g}$, where $a_{\rm g}$ is the radius of the gain medium aperture~\cite{a181221.01}. Then the beam radius at location $z$  can be computed by
\begin{equation}
w(z)=\dfrac{a_{\rm g}}{w_{00}(l+f)}\sqrt{-\dfrac{\lambda}{\pi\Im\left[1/q(z)\right]}}.
\label{equ:radius}
\end{equation}

Next, we calculate the power of the resonant beam. The light wave bounces in the SSLR circularly, where the resonant beam power varies during transferring inside the SSLR. According to the Rigrod analysis, we can calculate the power at each stage, as depicted in Fig~\ref{fig:model}(b). Here, we create an equivalent resonator for this calculation, and assume that the beam radius in the gain medium is a fixed value. The equivalent resonator only contains two equivalent mirrors and a gain medium, as shown in Fig.~\ref{fig:model}(b). The reflectivity of the equivalent mirrors are $\mathcal{R}_1$ and $\mathcal{R}_2$. Then, the power of the incident beam at the left-side equivalent mirror is obtained by~\cite{a181224.01}
\begin{equation}
P_4=\dfrac{\pi a_{\rm g}^2 I_{\rm s}}{(1+r_1/r_2)(1- r_2 r_1)}\left[\dfrac{l_{\rm g}\eta_{\rm c} P_{\rm in}}{I_{\rm s} V}  - \ln\dfrac{1}{r_2 r_1}\right],
\end{equation}
where $r_1\equiv\sqrt{\mathcal{R}_1}$ and $r_2\equiv\sqrt{\mathcal{R}_2}$ are the voltage reflection
coefficients of the equivalent mirrors, $I_{\rm s}$ is the saturation intensity, $l_{\rm g}$ is the gain medium thickness, $\eta_{\rm c}$ is the combined pumping efficiency, $V$ is the gain medium volume, and $P_{\rm in}$ is the electric power for driving the pump source.   $\mathcal{R}_1$ combines the power attenuation induced by the SHG and RR1. $\mathcal{R}_2$ combines the power attenuation induced by the gain medium, the air, and RR1.   They are expressed as
\begin{equation}
\left\{
\begin{array}{l}
\mathcal{R}_1=(1-\eta_{\rm SHG})^2\Gamma_{\rm SHG}^2 \Gamma_{\rm RR1}, \vspace{2ex}\\
\mathcal{R}_2=\Gamma_{\rm g}^2\Gamma_{\rm air}^2\Gamma_{\rm RR2}\Gamma_{\rm diff},
\end{array}
\right.
\end{equation}
where $\eta_{\rm SHG}$ is the SHG efficiency which represents the consumption of the resonant beam power in the SHG process; $\Gamma_{\rm SHG}$ is the transmittance of the SHG crystal without considering the loss induced by the SHG process; $\Gamma_{\rm RR1}$ and $\Gamma_{\rm RR2}$ are the loss factors of RR1 and RR2, respectively; $\Gamma_{\rm g}$ is the transmittance of the gain medium; $\Gamma_{\rm air}=\exp(-\alpha_{\rm air}d)$ is the transmittance of the air, where $\alpha_{\rm air}=0.0001$~m$^{-1}$ for clear air; $\Gamma_{\rm diff}$ is the diffraction loss factor which can be approximated by a closed-from formula presented in~\cite{MXiong2021}.  $\Gamma_{\rm RR1}$ is induced by the losses of passing through lens L1 and mirror M1. These losses mainly include the reflection of the surfaces of the lenses and the transmission of the mirrors, which are inevitable. Let $\Gamma_{\rm L1}$ denote the transmittance of L1, and $R_{\rm M1}$ denote the reflectivity of  M1, the loss factor of RR1 is expressed as
\begin{equation}
\Gamma_{\rm RR1}=\Gamma_{\rm L1}^2 R_{\rm M1}.
\end{equation}
Similarly, the loss factor of RR2 is expressed as
\begin{equation}
\Gamma_{\rm RR2}=\Gamma_{\rm L2}^2 R_{\rm M2},
\end{equation}
where $\Gamma_{\rm L2}$ is the transmittance of lens L2, and $R_{\rm M2}$ is the reflectivity of mirror M2. 

Now we have the expression of the resonant beam power outputting from the left-side surface of the gain medium. Since the SHG efficiency $\eta_{\rm SHG}$ is a function of $P_4$, we should solve the following equations to obtain the actual value of $P_4$:
\begin{equation}
\left\{\begin{array}{l}
P_4=\dfrac{\pi a_{\rm g}^2 I_{\rm s}}{(1+r_1/r_2)(1- r_2 r_1)}\left[\dfrac{l_{\rm g}\eta_{\rm c} P_{\rm in}}{I_{\rm s} V}  - \ln\dfrac{1}{r_2 r_1}\right], \vspace{2ex}\\
r_1=(1-\eta_{\rm SHG})\Gamma_{\rm SHG}\sqrt{\Gamma_{\rm RR1}}, \vspace{2ex}\\
r_2=\Gamma_{\rm g}\Gamma_{\rm air}\sqrt{\Gamma_{\rm RR2}\Gamma_{\rm diff}}, \vspace{2ex}\\
\eta_{\rm SHG}=  \dfrac{8 \pi^2 d_{\rm eff}^2 l_{\rm s}^2 }{ \varepsilon_0 c \lambda^2 n_0^3}\cdot \dfrac{2P_4}{\pi w^2(0)}, \\
\end{array}
\right.\label{equ:P4equs}
\end{equation}
where $\varepsilon_0$ is the vacuum permeability; $c$ is the speed of light; $d_{\rm eff}$, $l_{\rm s}$, and $n_0$ are the SHG crystal's efficient nonlinear coefficient, thickness, and refractive index, respectively~\cite{a181218.01,a200503.01}. Here, we assume the input of the SHG crystal is a plane wave. This is valid when $l_{\rm s}$ is smaller than the Rayleigh length $z_{\rm R}=\pi w_{00}^2(0)/\lambda$. Under this assumption, the beam radius is constant along the axis in the SHG crystal. $2P_4/[\pi w^2(0)]$ represents the resonant beam intensity that contains the leftward- and rightward-traveling wave at the SHG crystal. We also assume $\eta_{\rm SHG}$ is small, where the intensities of the leftward- and the rightward-traveling waves at the SHG crystal are the same. This condition can be satisfied because the communication demand for the carrier beam power is very small. There is no analytic solution for~\eqref{equ:P4equs}. Hence, we obtain the numerical solution using MATLAB.

\subsection{Communication Carrier Generation}
\begin{figure}[t]
\centering
\includegraphics[width=3.3in]{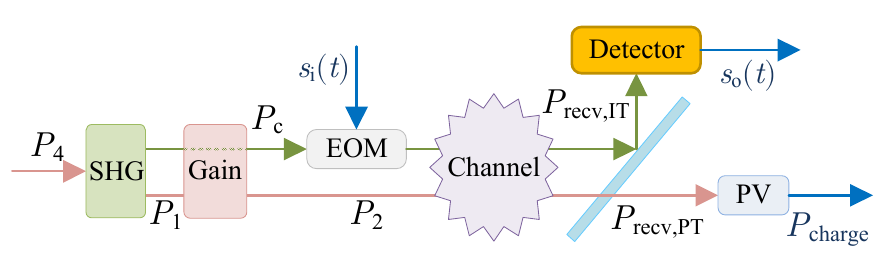}
\caption{Diagram of power and signal flows}
\label{fig:channel}
\end{figure}
 The SHG process is based on the nonlinear phenomenon of birefringent crystals, in which the crystal absorbs two fundamental photons and then generates a frequency-doubled photon. In our system, the resonant beam photons at frequency $\nu$ are the fundamental photons of the SHG process. Thus, the frequency-doubled photons have the frequency of $2\nu$. The communication carrier is the superposition of the frequency-doubled beams that generated from the leftward- and the rightward-traveling components of the resonant beam. Since the frequency of the communication carrier exhibits higher energy than that of the stimulated absorption  band of the gain medium atoms, the frequency-doubled beam can not interfere resonant beam generation. Thus, the resonant beam and the communication carrier transfer independently. At the receiver, the communication carrier is separated from the resonant beam by the frequency-selective coating of the dichroic mirror M5. This is easy to be realized, as the frequency difference between the communication carrier and the resonant beam is up to $282$~THz. Consequently, the effect of the dichroic mirror on the signal bandwidth can be neglected. With the aforementioned low-SHG-efficiency assumption, the carrier beam power is approximated as
\begin{equation}
P_{\rm c} =2\eta_{\rm SHG}P_4.
\end{equation}
Many materials, such as lithium niobate (LiNbO$_3$), potassium titanyl phosphate (KTP), and potassium dihydrogen
phosphate (KDP), can be employed for this purpose. Different materials exhibit different nonlinear coefficients and different phase matching angles. The incident angle of the input beam at the SHG crystal should be coincident with the  phase matching angle to reach the maximum conversion efficiency. Since the beams  between M1 and L1 are always perpendicular to these devices, a fixed beam incident angle at the SHG crystal can be satisfied. Hence, the crystal can be cut with proper orientation to ensure that the beam incident angle equals the phase matching angle. We choose LiNbO$_3$ as an example in this paper, since it can provides zero walk-off angle at $1064$-nm incident wavelength~\cite{a181218.01}. In fact, with the development of manufacturing, many advanced materials can be produced for this particular  application.

\subsection{Communication Channel and Achievable Rate}

As depicted in Fig.~\ref{fig:design} and Fig.~\ref{fig:channel}, the carrier is modulated by the EOM, transmitted through the channel, extracted by the dichroic mirror M5, and finally received by the detector. Along the transmission path, the signal is attenuated by the lenses, the mirrors, the gain medium, the EOM, and the air. These loss factors are caused by non-ideal reflection, non-ideal transmission, or absorption of impurity. Hence, the received optical power for information transfer is expressed as
\begin{equation}
P_{\rm recv,IT} =\Gamma_{\rm PD}\Gamma_{\rm L4}R_{\rm M5}^{(2\nu)} \Gamma_{\rm M2}^{(2\nu)}\Gamma_{\rm L2}\Gamma_{\rm air}\Gamma_{\rm g,EOM}\Gamma_{\rm L1}P_{\rm c},
\end{equation}
where $\Gamma_{\rm M2}^{(2\nu)}$, $\Gamma_{\rm L1}$, $\Gamma_{\rm L2}$, $\Gamma_{\rm L4}$,  $\Gamma_{\rm g,EOM}$, $\Gamma_{\rm air}$ are the transmittances of mirror M2, lens L1, lens L2, lens L4, the combined body of the gain medium and the EOM, and the air, respectively; $R_{\rm M5}^{(2\nu)}$ is the reflectivity of mirror M5; the superscript $(2\nu)$ represents that the operating frequency is $2\nu$; and $\Gamma_{\rm PD}$ represents the ratio of power received by the PD to the total incident power.  Generally, PD has a contractor on the front of the detecting area, which expands its receiving area $A_{\rm PD}$ to an effective signal collection area $A_{\rm eff}$. Let $A_{\rm o}$ denote the beam cross section area, the receiving ratio $\Gamma_{\rm PD}$ is expressed as:
\begin{equation}
	\Gamma_{\rm PD}=\left\{
	\begin{aligned}
		&A_{\rm eff}/A_{\rm o}, &A_{\rm eff}<A_{\rm o},\\
		&1, &A_{\rm eff}\geqslant A_{\rm o}.
	\end{aligned} 
	\right.
\end{equation}
The expression of $A_{\rm eff}$ is given as~\cite{a180730.03}
\begin{equation}
	A_{\rm eff}=\left\{
	\begin{aligned}
			&A_{\rm PD}T_{\rm s}(\psi)g(\psi)\cos\psi,&0\leqslant\psi\leqslant\Psi_{\rm c}\\
			&0,&\psi>\Psi_{\rm c}~~~~~~\\
	\end{aligned}
	\right.
\end{equation}  
where $g(\psi)$ is the concentrator gain, $T_{\rm s}(\psi)$ is the transmissivity of the concentrator surface, $\psi$ is the incidence angle with respect to the receiver axis, and $\Psi_{\rm c}$ is the semi-angle FOV of the concentrator. The expression of $g(\varphi)$ is given as~\cite{a180730.03}
\begin{equation}
	g(\psi)=\left\{
	\begin{aligned}
		&\dfrac{n_{\rm c}^2}{\sin^2\Psi_{\rm c}},&0\leqslant\psi\leqslant\Psi_{\rm c}\\
		&0,&\psi>\Psi_{\rm c}~~~~~~
	\end{aligned}
	\right.
\end{equation}  
where $n_{\rm c}$ is the internal refractive index of the concentrator. Since the branches of IT and PT are symmetric around M5, the received beam cross section area, $A_{\rm o}$, at PD is also approximated by~\eqref{equ:radius}.

From the received optical power to electric current, the conversion ratio is denoted by $\gamma$, which is usually called the PD's responsivity. Here we only consider the dominant noises, namely, the shot noise and the thermal noise. They are additive white Gaussian noise~(AWGN). The optical communication channel with intensity modulation is known as \emph{free-space optical
intensity channel}. Here, $P_{\rm recv,IT}$ represents the peak signal power that can be received. Generally, as a result of power consumption and safety consideration, there is an average power restriction for the intensity channel. The real average power depends on the modulation and encoding schemes. In our system, we assume the ratio of the average received power to the peak received power lies in the range  of $[0.5,1]$, as the energy resources and the safety are not significant restrictions in our system. Then, we find the achievable rate by computing the lower-bounded channel capacity; that is~\cite{a211104.01}
\begin{equation}
{R}_{\rm b}=\dfrac{1}{2}\log_2\left\{1+\dfrac{(\gamma P_{\rm recv,IT})^2}{2\pi e \sigma_{\rm n}^2}\right\},
\end{equation}
where $e$ is the nature constant. $\sigma_{\rm n}^2$ is the noise variance, which is obtained as~\cite{a200424.01}
\begin{equation}
	\sigma_{\rm n}^2=2\mathsf{e}(\gamma P_{\rm recv,IT} + I_{\rm bk}) B+\frac{4kT B}{R_{\rm IL}},
\end{equation}
where $\mathsf{e}$ is the electron charge, $k$ is Boltzmann constant, $I_{\rm bk}=5100~\mu$A  is the photon current induced by background radiation~\cite{a200427.02}, $B$ is the bandwidth, $T$ is the temperature in Kelvin, and $R_{\rm IL}$ is the load resistance. The optical devices on the transmission path of the modulated beam  are passive components. The light frequency of the communication carrier is up to $564$ THz. The signal bandwidth is much smaller than the carrier frequency. Thus, these passive optical devices have less effect to the signal bandwidth. Generally, the system bandwidth is limited by the EOM and the PD. There are several kinds of spatial EOMs, such as piezoelectric thin films~\cite{a200512.01} and multiple quantum well~(MQW) modulators~\cite{a200512.02}. Typical bandwidth of spatial EOMs is $800$~MHz to several gigahertzs. Some commercial PDs provide up to $100$-GHz bandwidth~\cite{a220303.02,100G2,a220303.03}. The actual bandwidths of the EOM and the PD depend on their incident areas. But the limitation of the incident area can be addressed by arrays structure~\cite{a200520.01,a211116.01}.

\subsection{ Charging Power at Receiver}

\begin{figure}[t]
	\centering
	\includegraphics[width=2.7in]{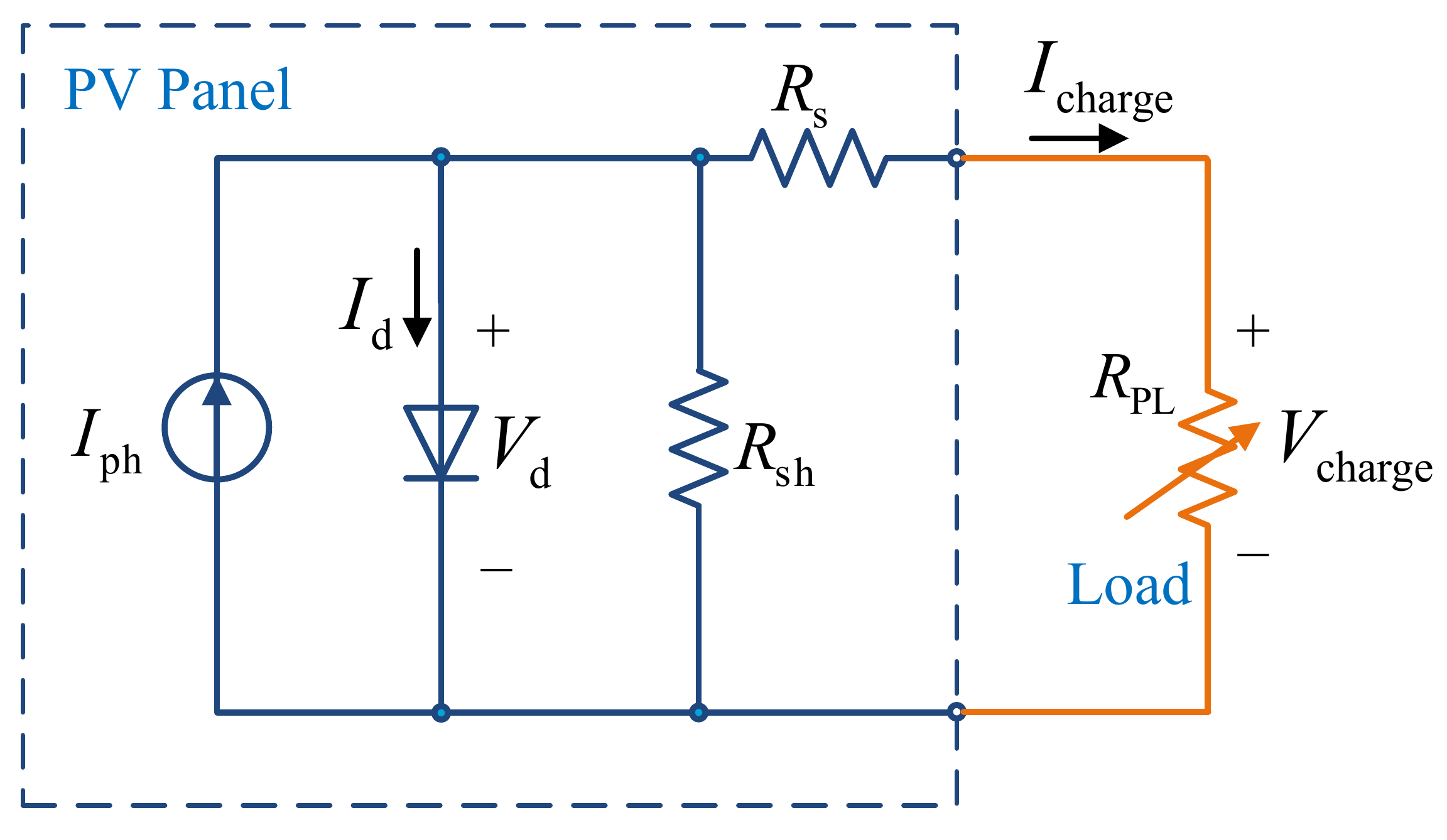}
	\caption{Equivalent circuit model for PV panel}
	\label{fig:PV}
\end{figure}

\begin{figure}[t]
	\centering
	\includegraphics[width=3.0in]{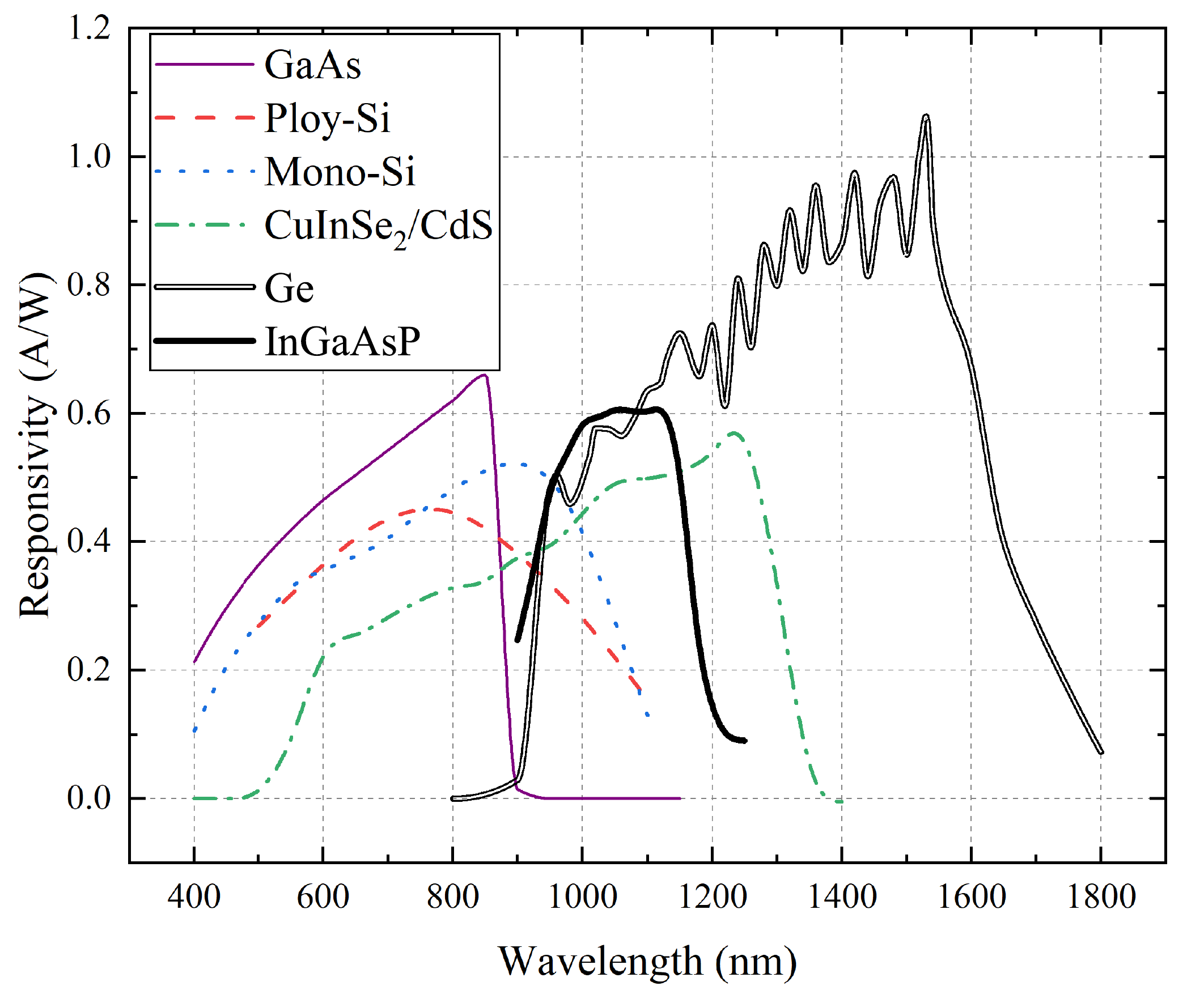}
	\caption{Spectral response of different photoelectric materials}
	\label{fig:responsivity}
\end{figure}

At the power transfer branch, the power $P_2$ of the traveling wave (at frequency $\nu$) outputting from the right-side surface of the gain medium (see Fig.~\ref{fig:model}) is transferred through the channel and received by the PV panel. Similar to the information transfer branch, the beam power experiences a series of attenuation induced by the air, the optical devices, and the PV. Therefore, the received optical power for 
wireless charging is expressed as
\begin{equation}
P_{\rm recv,PT} = \Gamma_{\rm PV}\Gamma_{\rm L3}  \Gamma_{\rm M5}^{(\nu)} \Gamma_{\rm M2}^{(\nu)}\Gamma_{\rm L2} \Gamma_{\rm air} P_2,
\end{equation}
where $\Gamma_{\rm L2}$, $\Gamma_{\rm L3}$ $\Gamma_{\rm M2}^{(\nu)}$, and $\Gamma_{\rm M5}^{(\nu)}$ are the transmittances of lens L2, lens L3, mirror M2, and mirror M5, respectively; the superscript $(\nu)$ represents that the specified operating frequency is $\nu$; and $\Gamma_{\rm PV}$ represents the percentage of the  received  power of the PV in the arrived PT beam power, not including the part that is reflected off the PV's receiving plane, as well as the part that is lost due to the limited size of the receiving area. All of these factors attenuate the beam power when the PT beam passes these components. As shown in Fig.~\ref{fig:model}(b), the power $P_2$ and $P_4$ come from the amplified $P_1$ and $P_3$, respectively; and the power $P_1$ and $P_3$ come from the attenuated $P_4$ and $P_2$, respectively. Thus, the following relations can be obtained:
\begin{equation}
\left\{
\begin{array}{l}
P_1=\mathcal{R}_1 P_4, \\
P_3=\mathcal{R}_2 P_2. \\
\end{array}
\right.
\label{equ:P1P3}
\end{equation}
We also know that the production of the leftward- and the rightward-traveling waves at arbitrary location  in the $z$-axis is a constant, namely~\cite{a181224.01}
\begin{equation}
P_1 P_4=P_2 P_3.
\label{equ:PP-Const}
\end{equation}
By combining these relations expressed in~\eqref{equ:P1P3} and~\eqref{equ:PP-Const}, we can obtain
\begin{align}
P_2&=\frac{r_1}{r_2}  P_4,
\end{align}
where $r_1$ and $r_2$ are roots of $\mathcal{R}_1$ and $\mathcal{R}_2$, respectively; and their expressions are given in~\eqref{equ:P4equs}. Consequently, the received optical power at the power transfer branch can be expressed as
\begin{equation}
P_{\rm recv,PT} = \Gamma_{\rm PV}\Gamma_{\rm L3}  \Gamma_{\rm M5}^{(\nu)} \Gamma_{\rm M2}^{(\nu)} \Gamma_{\rm L2} \Gamma_{\rm air} \left(\frac{r_1}{r_2} P_4\right).
\end{equation}

%

\begin{table} [t] 
	\caption{ Parameters of Laser Resonator}
	\renewcommand{\arraystretch}{1.2}
	\centering
	\begin{tabular}{ l l l}
		\hline
		\textbf{Parameter} & \textbf{Symbol} &  \textbf{Value} \\
		\hline
		Focal length of the lens & $f$ & $3$~cm\\
		Interval between the lens and mirror& $l$ & $3.015$~cm\\
		Saturation intensity of the gain medium & $I_{\rm s}$ &  $1.1976\times10^7$~W/m$^2$\\
		Resonant beam wavelength & $\lambda$ & $1064$ nm\\
		Radius of gain medium aperture  & $a_{\rm g}$ & $2$ mm \\
		Gain medium thickness & $l_{\rm g}$ & $1$ mm\\
		Combined pumping efficiency& $\eta_{\rm c}$ & $43.9\%$\\

		Efficient nonlinear coefficient & $d_{\rm eff}$ & $4.7$~pm/V\\
		Refractive index of SHG crystal&$n_0$ & 2.23 \\
		SHG crystal thickness & $l_{\rm s}$ & $0.4$ mm\\
		PV's responsivity&$\rho$ &$0.6$~A/W \\
		Reverse saturation current& $I_0$ & $0.32~\mu$A\\
		Shunt resistance& $R_{\rm sh}$ &$53.82$~$\Omega$ \\
		Series resistance & $R_{\rm s}$ & $37$~m$\Omega$\\
		Diode ideality factor& $n$ & $1.48$\\
		Number of cells in PV panel& $n_{\rm s}$ & $1$\\
		Temperature& $T$ & $298$~K \\
		PD's responsivity& $\gamma$ & $0.4$~A/W\\
		Load resistor & $R_{\rm IL}$& $10$~k$\Omega$\\
				\hline
	\end{tabular}
	\label{tab:paramReson}
\end{table}

The received optical power at the PV panel is converted into electricity for battery charging. As demonstrated in Fig.~\ref{fig:PV}, the PV panel can be equivalent to a current source in parallel with a diode. The current generated by the source is called the photo-current $I_{\rm ph}$, which is expressed as
\begin{equation}
I_{\rm ph}= \rho P_{\rm recv,PT},
\end{equation}
where $\rho$ is the optical-to-electrical conversion responsivity. However, for real devices, there are a shunt resistance $R_{\rm sh} $ and a series resistance $R_{\rm s}$  inside the PV model. Different photoelectronic material at different wavelength exhibits different responsivity. Fig.~\ref{fig:responsivity} shows the spectral response curves of different materials, including gallium arsenide (GaAs)~\cite{Mintairov2015GaAs}, polycrystalline silicon (Poly-Si)~\cite{Kumari1981PloySi}, monocrystalline silicon (Mono-Si)~\cite{Meyer2012MonoSi}, copper indium selenium/cadmium sulfide (CuInSe$_2$/CdS)~\cite{Kazmerski1976CuInSe2}, germanium (Ge)~\cite{Meusel2003Ge}, and indium gallium arsenide phosphide (InGaAsP)~\cite{a211116.03}.  According to  Kirchhoff's law, we can write down the following equations for this circuit model; that is~\cite{a190923.01,a180802.01}
\begin{equation}
\left\{
\begin{array}{lr}
I_{\rm charge}=I_{\rm ph}-I_{\rm d}-\dfrac{V_{\rm d}}{R_{\rm sh}}, \vspace{1ex}\\
I_{\rm d}=I_0 \left[ \exp\left({\dfrac{V_{\rm d}}{n_{\rm s}n V_{T}}}\right) -1 \right],  \vspace{1ex}\\
V_{\rm d}=I_{\rm charge}(R_{\rm PL}+R_{\rm s}), \vspace{1ex}\\
V_{T}=\dfrac{kT}{\mathsf{e}},
\end{array}
\right. \label{equ:PV}
\end{equation}
where $I_{\rm d}$ is the current passing through the diode; $V_{\rm d}$ is the voltage on the diode; $I_0$ is the reverse saturation current; $k$ is Boltzmann’s constant; $n$ is the diode ideality factor; and $n_{\rm s}$ is the number of cells connected in series in the PV panel. In this paper, we only consider single-cell PV panel, i. e., $n_{\rm s}=1$.  We also obtain the numeric solution of ~\eqref{equ:PV}, as it has no analytical solution.

For a certain $P_{\rm recv,PT}$, the PV panel performs like a constant-current source unless the output voltage $V_{\rm charge}$ approaches the open-circuit voltage. The PV can achieve a maximum power point~(MPP) if we chose a proper charging voltage. Generally, this work is implemented by a maximum power point tracking~(MPPT) device connected to the battery. To obtain the maximum charge power $\hat{P}_{\rm charge}$, we  solve the following problem
\begin{align}
(\mathbf{P1}):~~&\hat{P}_{\rm charge} =\max\limits_{\{V_{\rm charge}\}}  I_{\rm charge} V_{\rm charge}, \vspace{2ex}\\
&\mbox{ s.t.}~~\left\{
\begin{array}{l}
R_{\rm PL}=\dfrac{V_{\rm charge}}{I_{\rm charge}},\vspace{1ex}\\
0\leqslant V_{\rm charge}~\leqslant V_{\rm oc},
\end{array}\right.\label{equ:Pchg-max-st}
\end{align}
In~\eqref{equ:Pchg-max-st}, $R_{\rm PL}$ is the equivalent load resistance of the circuit connected to the PV panel. It is the MPPT device that adjusts the load resistance automatically to find the MPP. $V_{\rm oc}$ is called the open-circuit voltage, which is the output voltage of the PV panel when no load circuit is connected.

\section{Numerical Results}
\label{sec:result}

\subsection{Parameters Setting}

Unless otherwise specified, the parameters of our system are listed in Table~\ref{tab:paramReson}. In this paper, an Nd:YVO$_4$ crystal is used as the gain medium. Hence, the wavelength of the resonant beam  $\lambda=1064$~nm, and the  information carrier is the $532$-nm frequency-doubled beam. Nd:YVO$_4$ is an efficient crystal for diode-pump solid-state lasers. It has large stimulated emission cross-section at lasing wavelength and high absorption coefficient at pumping wavelength. Thus, the crystal can be cut into a thin disk to satisfy the FOV requirement of our system. The parameters of the gain medium material are obtained from~\cite{a181218.01},  including the saturation intensity $I_{\rm s}$, the stimulated wavelength $\lambda$, and the combined pump efficiency $\eta_{\rm c}$. 

The parameters of the SHG material are also obtained from~\cite{a181218.01}, including the efficient nonlinear coefficient $d_{\rm eff}$ and the refractive index $n_0$. The shape parameters of the gain medium and the SHG medium  are customized in this paper according to the requirements. In our SSLR-based wireless charging experiment, the diameters of the lenses and mirrors are $14$~mm~\cite{RBCexperiment}. Since the apertures of the lenses and mirrors are much bigger than the gain medium, their effects can be neglected in calculation. These lenses and mirrors are made of N-BK7 glasses. Here, the focal length of each lens is set as $3$~cm. The space interval between the lens and the mirror in each TCR is set as $3.015$~cm. Then, $f_{\rm RR}$ is calculated as $3$~m, and thus, the upper boundary of the  distance is $12$ m. 

For devices through which the beams pass, including the lenses, the SHG crystal, and the dichroic mirror, unwanted reflection at their input and output surfaces exists even if these devices are with the best production.  We set the transmittance at each AR surface as $99.5\%$ which is a typical value of commercial products. Therefore, we set $\{\Gamma_{\rm L1}, \Gamma_{\rm L2}, \Gamma_{\rm L3},  \Gamma_{\rm L4}, \Gamma_{\rm SHG}, \Gamma_{\rm M2}^{(2\nu)}, \Gamma_{\rm M5}^{(\nu)}\} = 99\%$. Also the receiving surfaces of PD and PV have reflectivity of $0.5\%$; hence, we set $\{T_{\rm s}, \Gamma_{\rm PV}\}=99.5\%$. The receiving area of such high-speed PD should be $A_{\rm PD}=0.4\times0.4~\mbox{mm}^2$~\cite{a211116.01}. We set the semi-angle FOV of the PD as $\Psi_{\rm c}=30^\circ$ and the refractive index of concentrator as $n_{\rm c}=1.5$~\cite{a220305.01}. The incidence angle $\psi=0$. HR mirrors/coatings also have inevitable transmittance; hence, we set their real reflectivity $\{R_{\rm M1}, R_{\rm M5}^{(2\nu)}\}=99.5\%$. As the resonant beam passes  the gain medium, it meets the front AR surface twice and the back HR surface once. Therefore, we set $\Gamma_{\rm g}=98.51\%$.  The frequency-doubled beam meets interfaces five time as it passes the combined body of the gain medium and the EOM. Hence, we set $\Gamma_{\rm g,EOM}=97.52\%$. Besides, we set the bandwidth $B=800$~MHz which is a feasible value of electro-absorption modulators~\cite{a200520.01}. From the spectral response curves shown in Fig.~\ref{fig:responsivity}, we found that InGaAsP performs well at $1064$~nm (responsivity is about $0.6$~A/W), while GaAs has a good responsivity ($0.4$~A/W) at $532$~nm. Therefore, we choose InGaAsP and GaAs as the materials of PV and PD, respectively. Other manufacturing-dependent parameters for PD and PV are obtained from~\cite{a200427.02} and \cite{a210622.02}, respectively.

\subsection{Performance Evaluation}
\begin{figure}[t]
	\centering
	\includegraphics[width=3.2in]{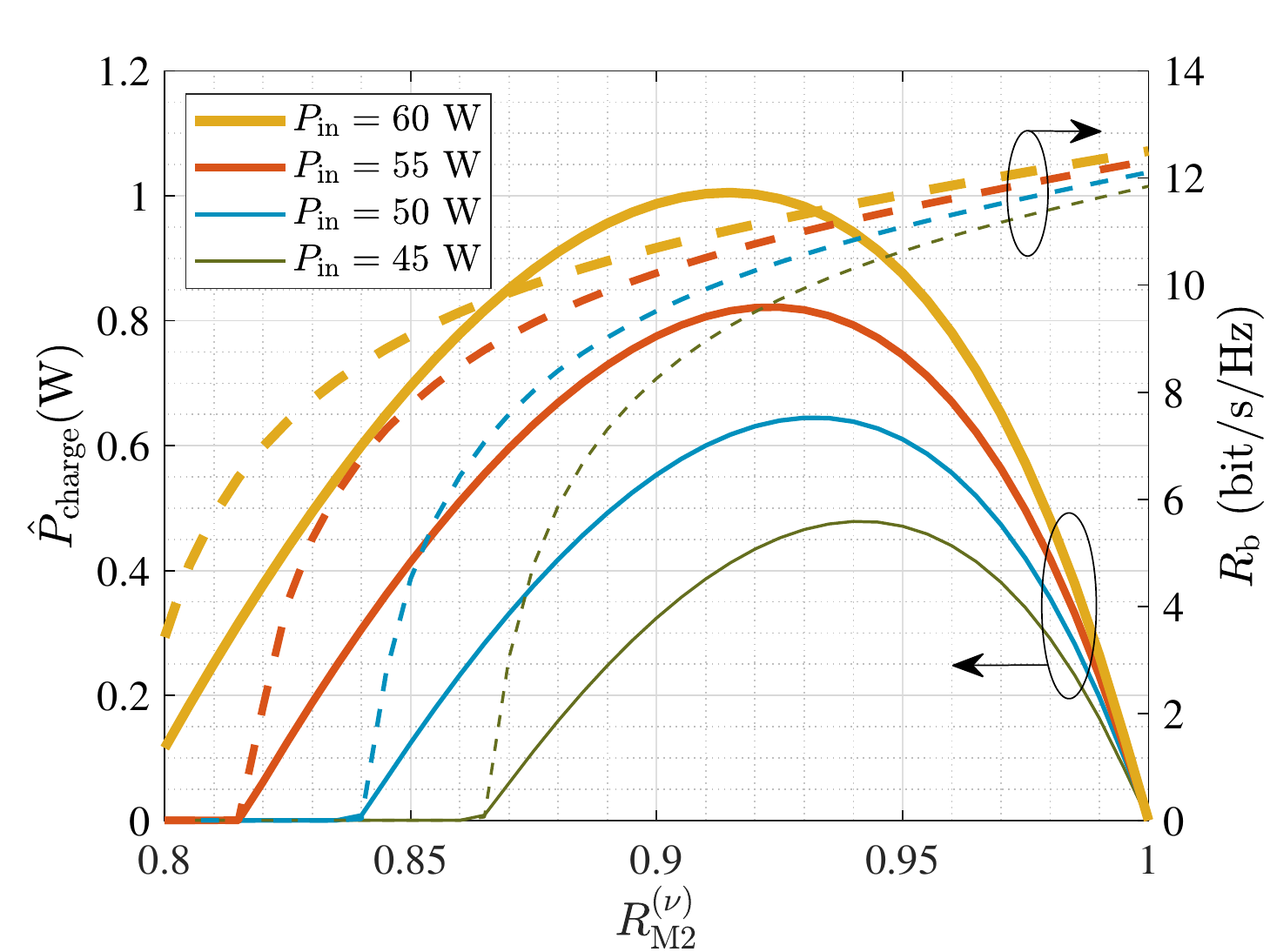}
	\caption{Maximum charging power $\hat{P}_{\rm charge}$ and achievable rate $R_{\rm b}$ \textit{vs.} reflectivity of mirror M2 $R_{\rm M2}^{(\nu)}$ (solid lines: $\hat{P}_{\rm charge}$; dashed lines: $R_{\rm b}$; $P_{\rm in}$: source power; distance $d=6$~m; SHG crystal thickness $l_{\rm s}=0.4$~mm)}
	\label{fig:PC-vs-RM2}
\end{figure}

Fig.~\ref{fig:PC-vs-RM2} demonstrates the maximum charging power $\hat{P}_{\rm charge}$ achieved by the MPPT device as a function of the reflectivity $R_{\rm M2}$ of the mirror M2, for different source power $P_{\rm in}$. As $R_{\rm M2}^{(\nu)}$ increases, $\hat{P}_{\rm charge}$ gradually increases to a maximum point and then decreases to $0$. The maximum value of $\hat{P}_{\rm charge}$ can be obtained with a specific $R_{\rm M2}^{(\nu)}$.  Moreover, as $P_{\rm in}$ grows, the optimal value of $R_{\rm M2}^{(\nu)}$ appears slightly decreasing. We also evaluate the achievable rate $R_{\rm b}$. As depicted in Fig.~\ref{fig:PC-vs-RM2}, with the increase of $R_{\rm M2}^{(\nu)}$,  $R_{\rm b}$ increases rapidly at first; and when $R_{\rm M2}^{(\nu)}$ reaches a large value, the change of $R_{\rm b}$ gradually slows down. When $P_{\rm in} = 60$~W and $R_{\rm M2}^{(\nu)} = 91.5\%$, $\hat{P}_{\rm charge}$ reaches the maximum value, i.e.,  $1.05$~W;  in this case, $R_{\rm b}=11.03$~bit/s/Hz.

\begin{figure}[t]
	\centering
	\includegraphics[width=3.2in]{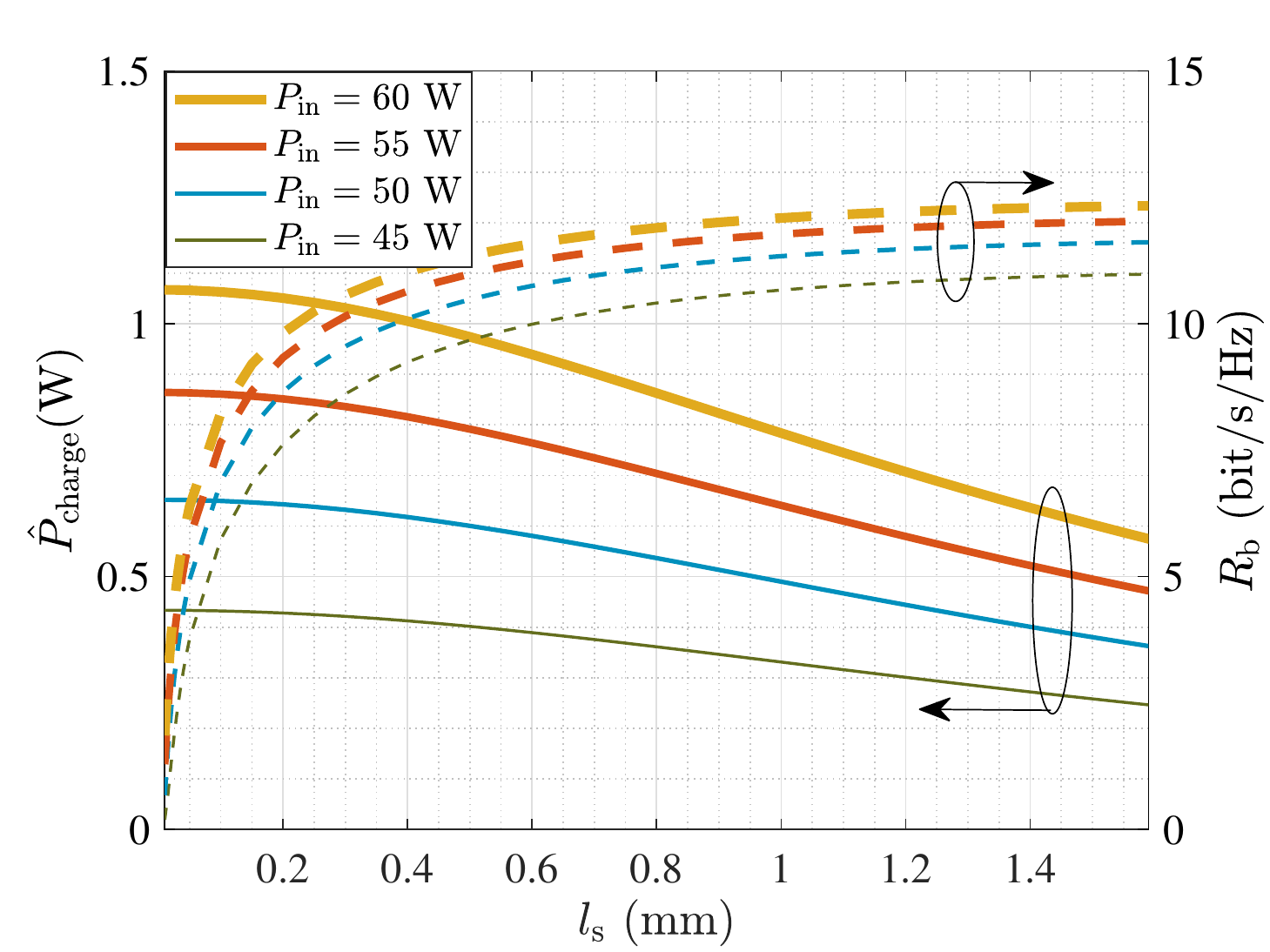}
	\caption{Maximum charging power $\hat{P}_{\rm charge}$ and achievable rate $R_{\rm b}$ \textit{vs.} SHG crystal thickness $l_{\rm s}$ (solid lines: $\hat{P}_{\rm charge}$; dashed lines: $R_{\rm b}$; $P_{\rm in}$: source power; distance $d=6$~m; M2's reflectivity $R_{\rm M2}^{(\nu)}=91.5\%$)}
	\label{fig:PC-vs-ls}
\end{figure}

\begin{figure}[t]
	\centering
	\includegraphics[width=3.2in]{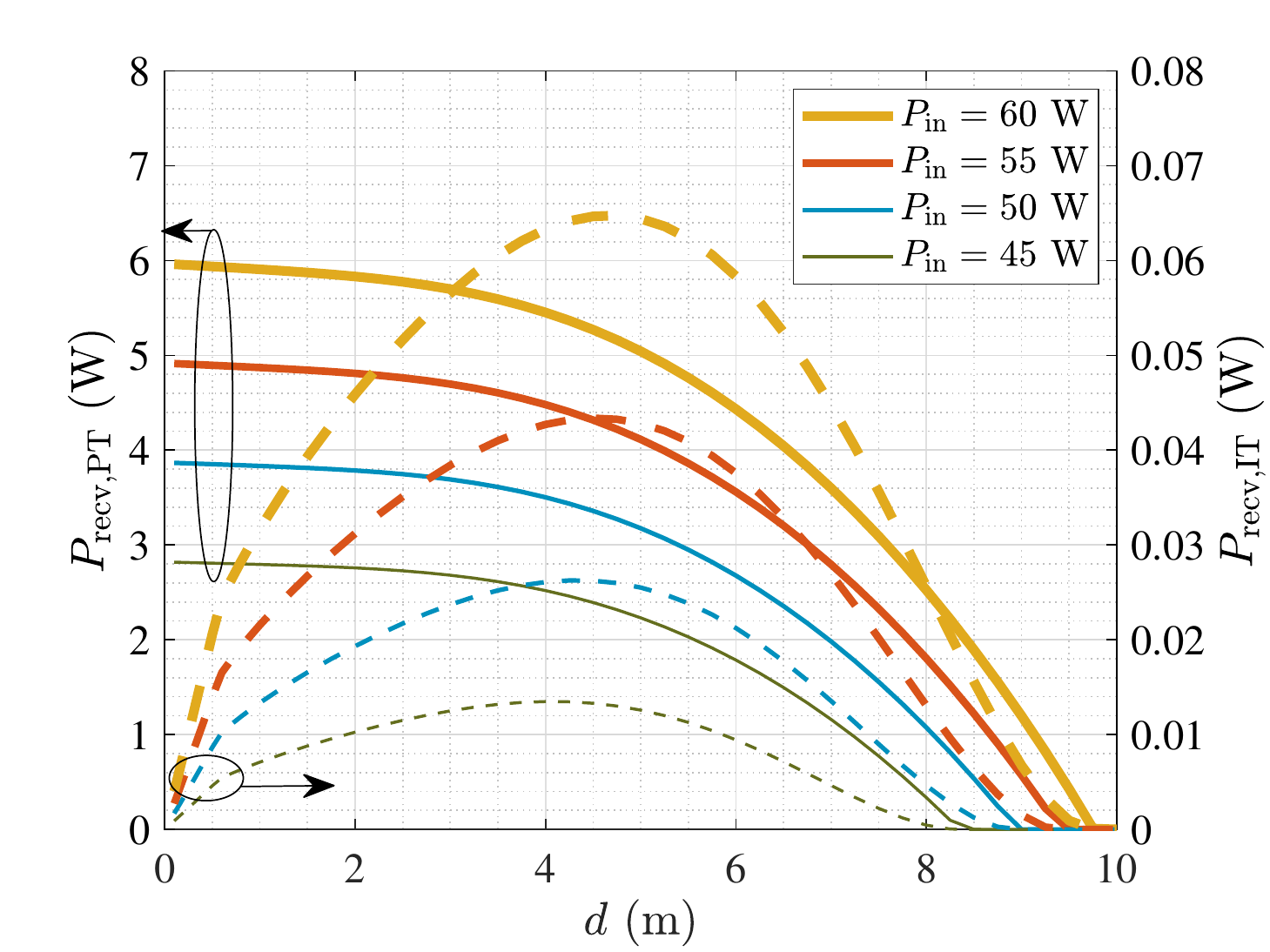}
	\caption{Received PT beam power $P_{\rm recv,PT}$ and IT beam power $P_{\rm recv,IT}$ \textit{vs.} distance $d$ (solid lines: $P_{\rm recv,PT}$; dashed lines: $P_{\rm recv,IT}$; $P_{\rm in}$: source power; M2's reflectivity $R_{\rm M2}^{(\nu)}=91.5\%$; SHG crystal thickness $l_{\rm s}=0.4$~mm)}
	\label{fig:PPT-PIT-vs-d}
\end{figure}

\begin{figure}[t]
	\centering
	\includegraphics[width=3.2in]{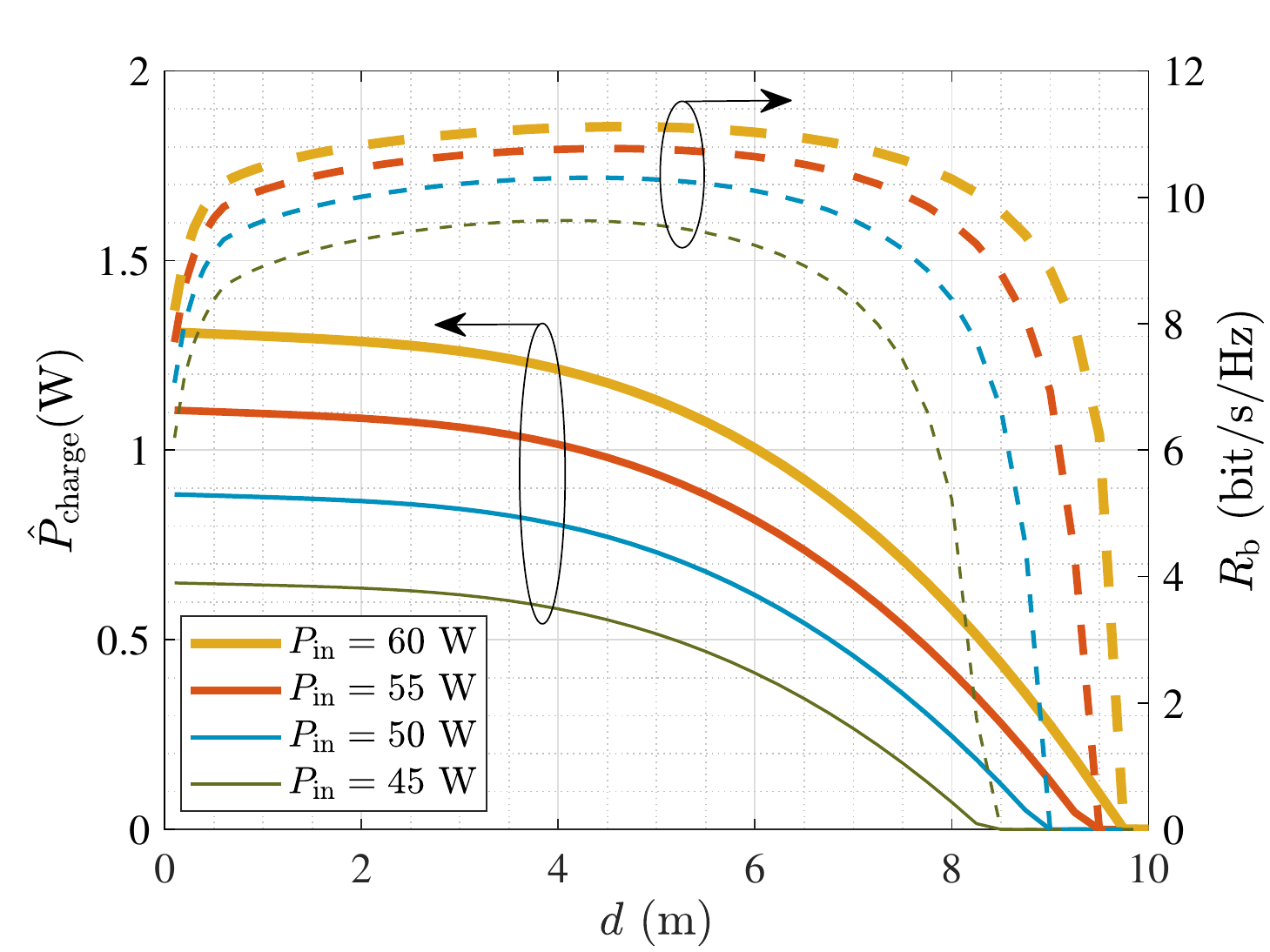}
	\caption{Maximum charging power $\hat{P}_{\rm charge}$ and achievable rate $R_{\rm b}$ \textit{vs.} distance $d$ (solid lines: $\hat{P}_{\rm charge}$; dashed lines: $R_{\rm b}$; $P_{\rm in}$: source power; M2's reflectivity $R_{\rm M2}^{(\nu)}=91.5\%$; SHG crystal thickness $l_{\rm s}=0.4$~mm)}
	\label{fig:PC-vs-d}
\end{figure}

Fig.~\ref{fig:PC-vs-ls} depicts the maximum charging power $\hat{P}_{\rm charge}$ and the achievable rate $R_{\rm b}$ as a function of the SHG crystal thickness $l_{\rm s}$, for different source power $P_{\rm in}$. According to~\eqref{equ:P4equs}, the thickness of the SHG crystal $l_{\rm s}$ affects the SHG efficiency $\eta_{\rm SHG}$. As shown in Fig.~\ref{fig:PC-vs-ls}, as $l_{\rm s}$ increases, $R_{\rm b}$ increases rapidly at first and then gradually approaches the highest point. This is because $\eta_{\rm SHG}$ increases with $l_{\rm s}$, and thus more frequency-doubled photons are generated. Conversely, the intra-cavity resonant beam power decreases gradually with the increase of $\eta_{\rm SHG}$, leading to the decrease of $\hat{P}_{\rm charge}$.

We analyzed the received PT beam power $P_{\rm recv,PT}$ and IT beam power $P_{\rm recv,IT}$ at difference transmission distance $d$, as shown in Fig.~10. $P_{\rm recv,PT}$ is at watt-level, while $P_{\rm recv,IT}$ has several tens of milliwatts. With the increase of the  distance, $P_{\rm recv,PT}$ decreases with a slow speed at first, and the decrease speed increases gradually. However, with the increase of the distance, $P_{\rm recv, IT}$ increases at first. This phenomenon can be explained by the decrease of the beam radius at the SHG crystal with the growth of distance, which results in the increasing resonant beam intensity and SHG efficiency. When $d>5$~m, $P_{\rm recv,IT}$ turns to decrease due to the increase of the diffraction loss.  In addition, we analyzed the changes of the maximum charging power $\hat{P}_{\rm charge}$ and the achievable rate $R_{\rm b}$ with the distance $d$, as shown in Fig.~\ref{fig:PC-vs-d}. As $d$ increases, $\hat{P}_{\rm charge}$ decreases gradually. However, the achievable rate $R_{\rm b}$ grows smoothly over most of the transmission range. When $P_{\rm in}=60$~W and the distance is less than $6$~m, $\hat{P}_{\rm charge}$ keeps above $1$~W, which is suitable for most indoor charging scenarios. For example, a typical mobile phone which is charged with $5$~V and $1$ A (i.e., the charge power is $5$~W) can be fully charged from $0\%$ to $100\%$ for $1$ hour. When we use the RB-SWIPT system, the phone can be fully charged within $5$~hours. Although the charging time is extended, the phone can be charged in use or in idle time, without the disturbing of cables. Similarly, $R_{\rm b}$ keeps above $10$ bit/s/Hz within a  large range of distance. Only when the distance is less than $0.2$~m or greater than $8$~m does $R_{\rm b}$ decrease rapidly to zero.

\begin{figure}[t]
	\centering
	\includegraphics[width=3.2in]{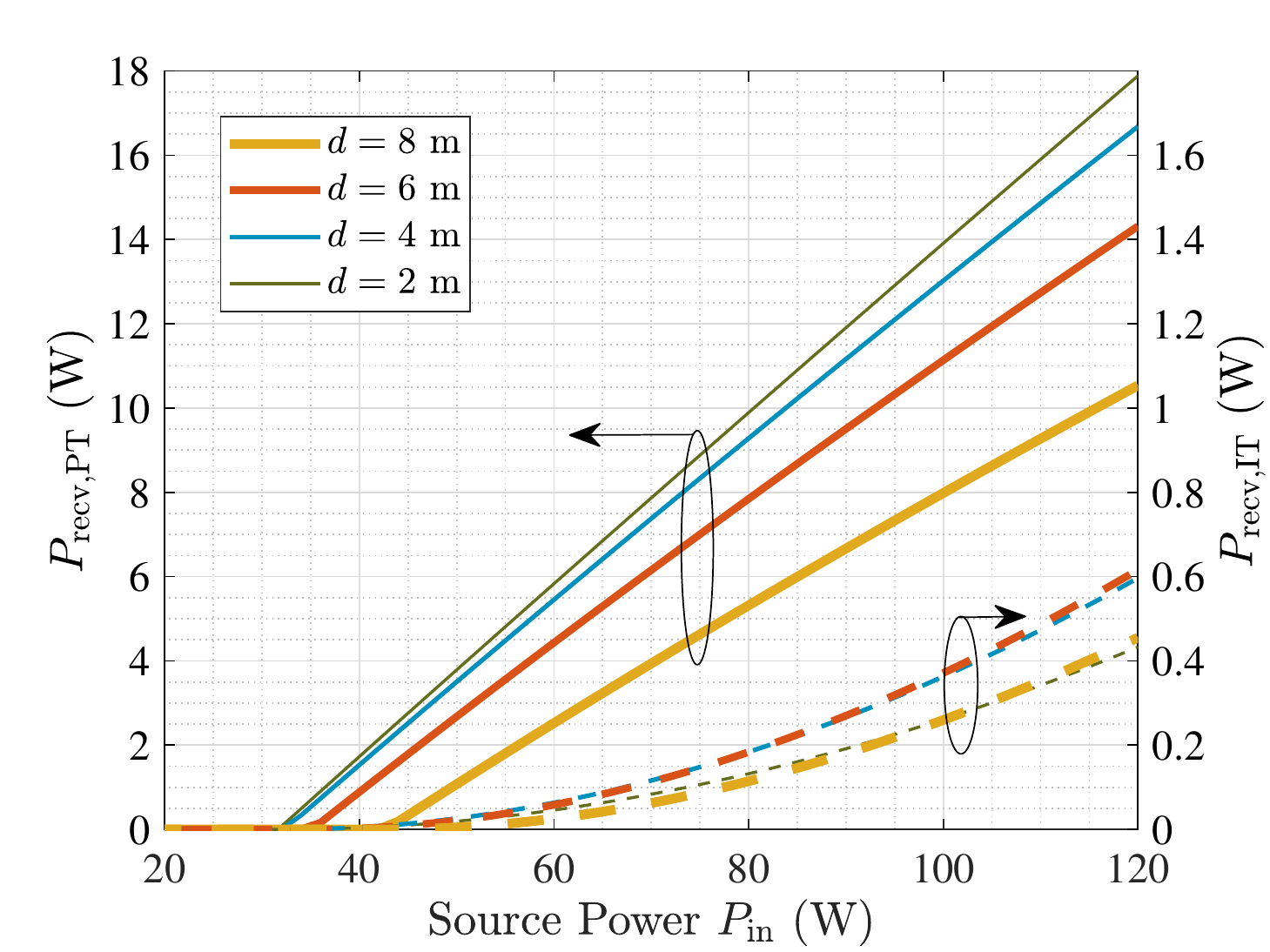}
	\caption{Received PT beam power $P_{\rm recv,PT}$ and IT beam power $P_{\rm recv,IT}$ \textit{vs.} source power $P_{\rm in}$ (solid lines: $P_{\rm recv,PT}$; dashed lines: $P_{\rm recv,IT}$; $d$: distance; M2's reflectivity $R_{\rm M2}^{(\nu)}=91.5\%$; SHG crystal thickness $l_{\rm s}=0.4$~mm)}
	\label{fig:PPT-PIT-vs-Pin}
\end{figure}

\begin{figure}[t]
	\centering
	\includegraphics[width=3.2in]{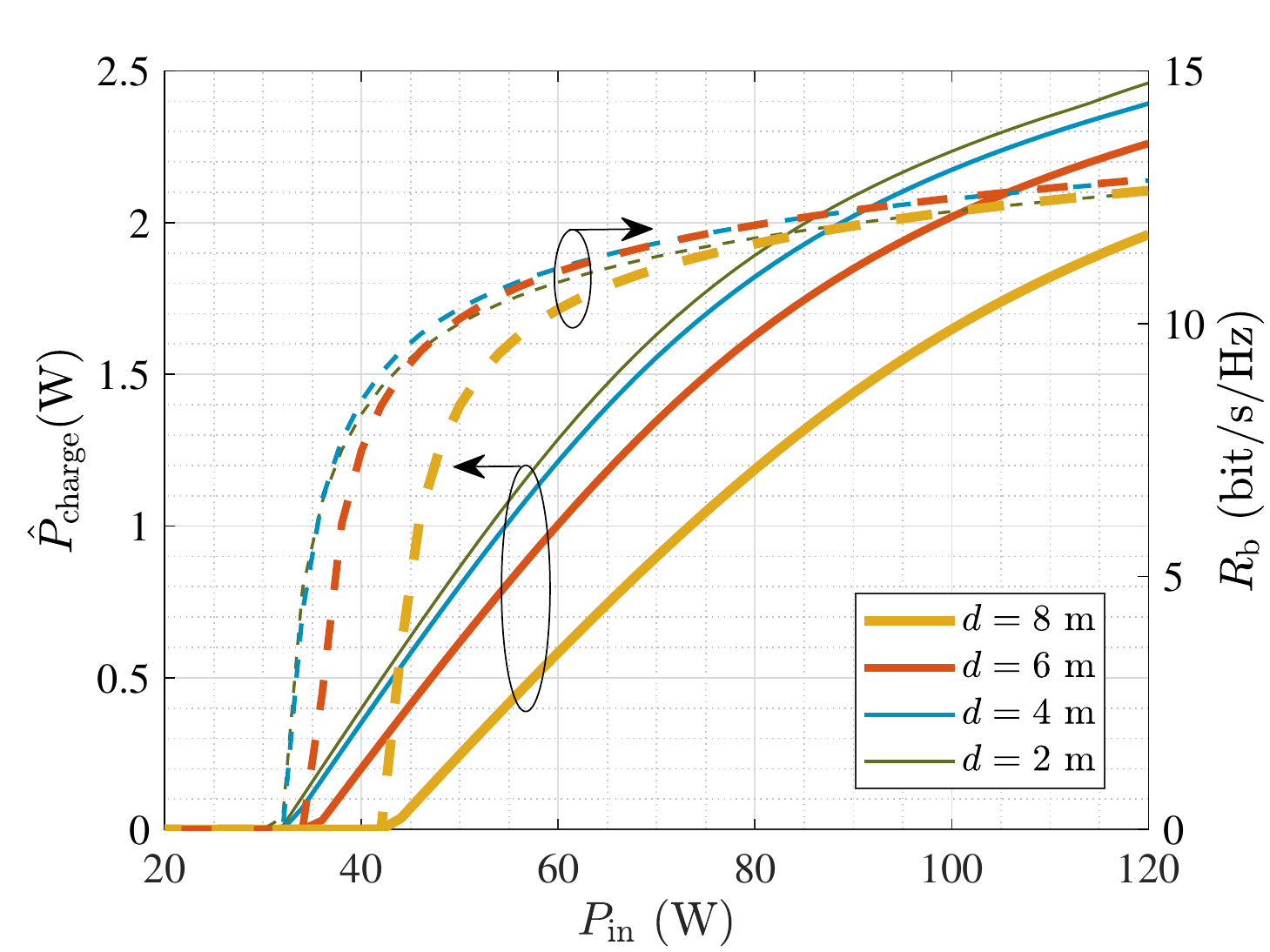}
	\caption{Maximum charging power $\hat{P}_{\rm charge}$ and achievable rate $R_{\rm b}$ \textit{vs.} source power $P_{\rm in}$ (solid lines: $\hat{P}_{\rm charge}$; dashed lines: $R_{\rm b}$; $d$: distance; M2's reflectivity $R_{\rm M2}^{(\nu)}=92\%$; SHG crystal thickness $l_{\rm s}=0.4$~mm)}
	\label{fig:PC-vs-Pin}
\end{figure}

As depicted in Fig.~\ref{fig:PPT-PIT-vs-Pin}, the relation between the received PT beam power $P_{\rm recv, PT}$ and the pump source power $P_{\rm in}$ is linear. However, as $P_{\rm in}$ grows, the increase of $P_{\rm recv, IT}$ accelerates. This is because the resonant beam intensity at the SHG crystal is also increased, which leads to the increase of the SHG efficiency. There is a threshold of $P_{\rm in}$ for each case. If $P_{\rm in}$ is below the threshold, the resonance can not be established, and thus the received powers are zeros.
As shown in Fig.~\ref{fig:PC-vs-Pin}, the maximum charging power $\hat{P}_{\rm charge}$ and the achievable rate $R_{\rm b}$ also depend on $P_{\rm in}$.  After $P_{\rm in}$ exceeds the threshold, $\hat{P}_{\rm charge}$ and $R_{\rm b}$ start to increase. $R_{\rm b}$ grows rapidly as $P_{\rm in}$ exceeds the threshold, and quickly reaches a slow growth rate. For a large range of $P_{\rm in}$, rate $R_{\rm b}$ can achieve over $10$~bit/s/Hz. As $P_{\rm in}$ continues to increase, the growth rate of $\hat{P}_{\rm charge}$ reduces gradually. This phenomenon can be explained by the decreased PV efficiency with the increase of the incident optical power. From~Fig.~\ref{fig:PC-vs-Pin}, we can also observe that, by choosing an appropriate $P_{\rm in}$, the RB-SWIPT system can achieve watt-level battery charging power and over $8$-m distance.
The above performance can support most indoor mobile devices with permanent battery life and high-quality multimedia stream transmission.

\section{Discussion}
\label{sec:discu}
\subsection{Safety}
\begin{figure}[t]
	\centering
	\includegraphics[width=3.2in]{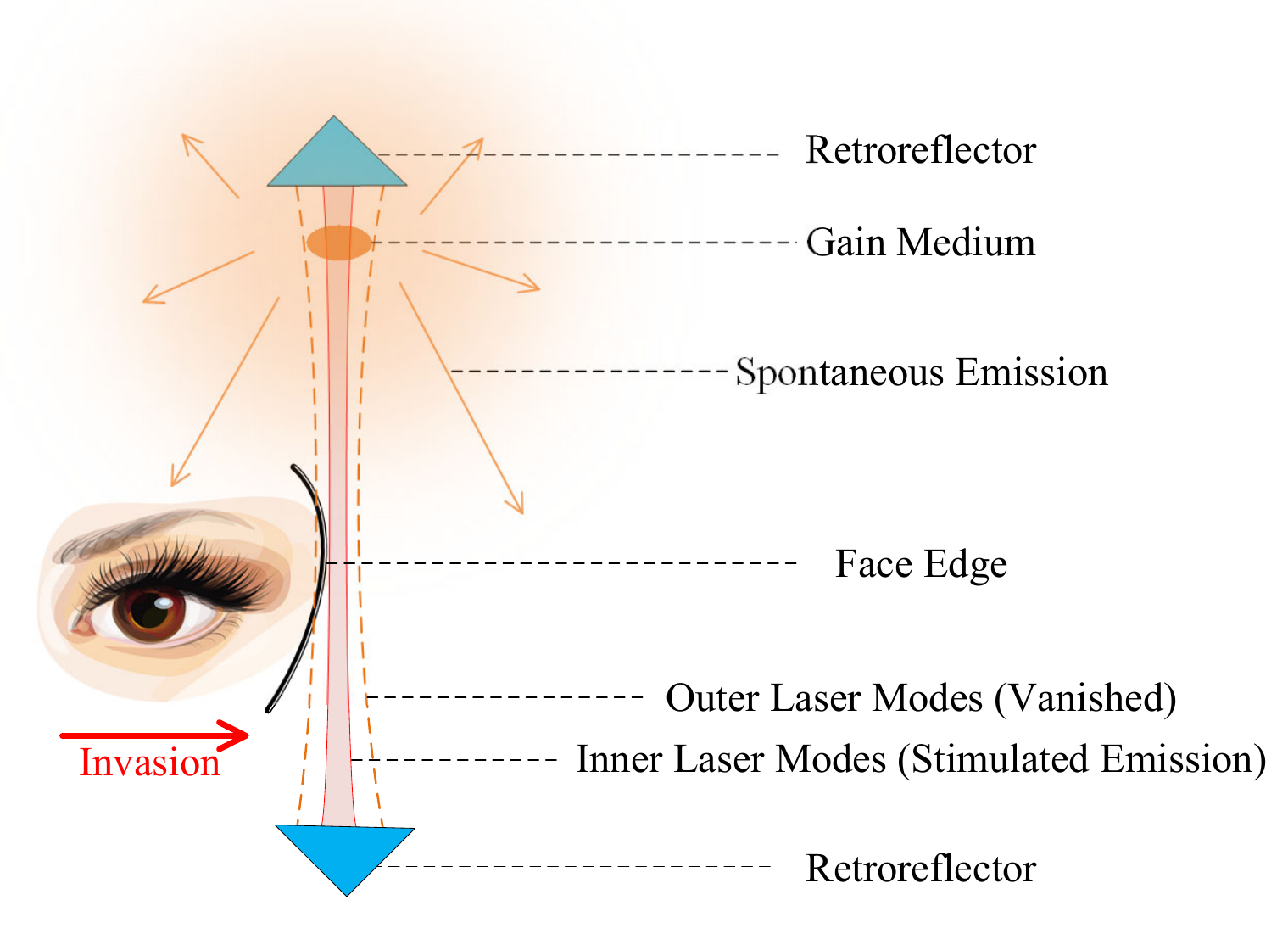}
	\caption{Safety analysis for human eyes invasion (considering both spontaneous emission and stimulated emission)}
	\label{fig:safety}
\end{figure}

Both skin safety and eye safety should be considered when implementing a resonant beam system. Laser products have been classified into seven classes according to the international electrotechnical commission (IEC) standards IEC 60825-1 and IEC 60825-12~\cite{a211117.01}. For eye safety under long-time exposure, the product should satisfy the maximum exposure energy~(MPE) restriction of Class 1 and Class 1M. This depends on not only the power density on the eyes but also the wavelength and the size of the source, i.e., it depends on whether the source in the eye's view is a point source or an extended source. As shown in Fig.~\ref{fig:safety}, there are spontaneous emission and stimulated emission in the resonant beam system. The spontaneous emission is omnidirectional. When the resonance has not been initiated or has been interrupted by obstacles, many excited atoms (have absorbed the pump power) in the gain medium lie at the higher energy level, and they transit to the lower energy level by emitting photons spontaneously. The stimulated emission (i.e., the resonant beam) is directional, as it occurs relying on the intra-cavity resonance and the optical amplification. 

\begin{figure}[t]
	\centering
	\includegraphics[width=3.2in]{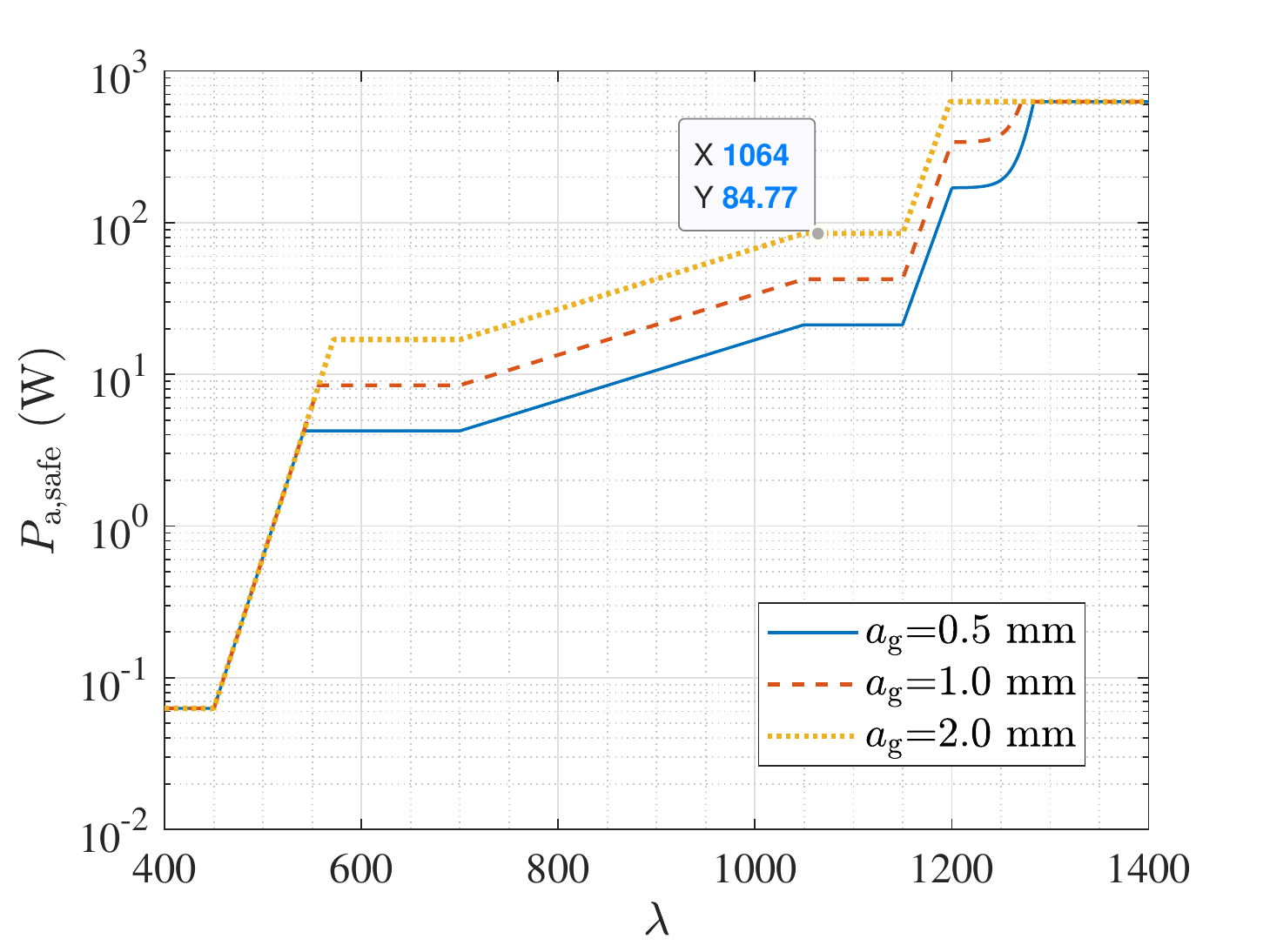}
	\caption{Maximum permissible  absorbed power of the gain medium $P_{\rm a,safe}$ under eye-safety restriction varies with wavelength $\lambda$}
	\label{fig:safepower}
\end{figure}

Now, we discuss the safety of these two cases. Since the spontaneous emission is omnidirectional and the gain medium diameter $a_{\rm g} \gg 0.2$~mm, the gain medium is classified as an extended source, as defined in IEC~60825~\cite{IEC60825.1}. According to the standard, the MPE of Class I extend source with radius $a_{\rm g}=2~\mbox{mm}$ (angular of subtense $\alpha=40$~mrad) and  $\lambda=1064$~nm is computed as $0.1349$~W/cm$^2$. If the input source power~$P_{\rm in}$ to the pump diode is $60$~W, the actual power absorbed by the gain medium, $P_{\rm a}$, should be obtained by multiplying~$P_{\rm in}$ and a series of efficiencies, including the pump source efficiency $\eta_{\rm P}=75\%$, the pump laser transmission efficiency $\eta_{\rm t}=99\%$, and the gain medium absorption efficiency $\eta_{\rm a}=91\%$~\cite{a181218.01}; that is $40.5$~W. Since there is an HR coating behind the gain medium, the real spontaneous emission power intensity at the most restrictive measurement distance $d_{\rm e}=10$~cm is approximated as $2\eta_{\rm P}\eta_{\rm t}\eta_{\rm a}P_{\rm in}/(4\pi d_{\rm e}^2)=0.0645$~W/cm$^2$, which is under the MPE value. Different wavelengths $\lambda$ have different MPE values. The maximum permissible absorbed powers $P_{\rm a,safe}$ under different $\lambda$ and $a_{\rm g}$ are depicted in Fig.~\ref{fig:safepower}. We can see that $P_{\rm a,safe}=84.77$~W under the parameters specified in this paper. Then the maximum permissible source power is calculated as $P_{\rm in, safe}=P_{\rm a,safe}/(\eta_{\rm P}\eta_{\rm t}\eta_{\rm a})=125.46$~W. To verify the eye safety of real products, rigorous measurements and ethical animal experiments are still needed. Hence, it is necessary to wear a protective glasses for working wavelengths during the laser experiment.

In fact, when the stimulated emission becomes dominant, the power of spontaneous emission reduces largely. This gives us the opportunity to reduce the spontaneous emission by adjusting the pump progress. Namely, we can use a low source power to generate the resonant beam and detect the resonance by a sensor. Only when the resonance is detected do we further increase the source power. Besides, using wavelengths greater than $1400$~nm can reduce the risk of retina injury as these wavelengths are absorbed by the vitreous humor in eyes. 

The stimulated emission can not reach the eyes, as it is interrupted by the obstruction of  human face. The resonant beam is very thin, generally with millimeter-level radius, and it consists of many transverse modes. From high-order to low-order modes, the mode radius decreases. Those modes oscillate in the cavity independently. When a human face is entering the resonant beam, the mode with the greatest radius vanish first due to the increase of the diffraction loss. The power is then reallocated to the inner oscillating modes. This energy transfer process continues as the human face invasion process progresses. The interval between the edges of two adjacent modes are less than $1$~mm, and the ratio of the energy outside the mode radius to the total mode energy is less than $1/e^2$. Even though the diffraction energy of the outer modes is released on the face edge, the power reaching the eyes is negligible. In addition, as the overlap between the resonant beam and the gain medium becomes small due to the vanishing of outer modes, the power used for stimulated emission is decreased, leading to the decrease of the resonant beam power. 

Quantitatively analysis with light-field simulation on the diffraction energy received by an invading object is demonstrated in~\cite{WFang2021}, which verified the skin safety of the resonant beam system under the given parameters. We also experimentally verified the skin safety of the resonant beam system driven by $40$-W pump laser, as presented in~\cite{RBCexperiment}. Parameter optimization can help reduce the power exposure during intrusion. For instance, improving the gain  of the gain medium and correspondingly  decreasing the reflectivity, $R_{\rm M2}^{(\nu)}$, of the output mirror M2 can reduce the intra-cavity beam power and meanwhile keep the output power unchanged. This principle may motivate new designs for the resonant beam system, including the SSLR structure, the gain medium material, and the pump module structure. 

\subsection{Energy Efficiency}
	Recently, the energy efficiency in this paper is around $2\%$ (only consider the PT branch; $P_{\rm in}=80$~W; $d=6$~m). The power losses occur in several steps: \ding{172} electro-optical conversion of pump laser diode, \ding{173} pump laser transmission, \ding{174} gain medium absorption, \ding{175} spontaneous emission and \ding{176} nonradiative transition of excited atoms, \ding{177} diffraction loss during beam propagation, \ding{178} passing the intra-cavity optical devices, \ding{179} over-the-air  transmission, and \ding{180} photovoltaic conversion. In this work, the parameters of the pump source module and  the gain medium are obtained from the experiment conducted in~\cite{a181218.01}. However, with the development of technologies, the laser devices have been constantly improved. For example, if the gain medium is pumped directly with electricity, referring to vertical-cavity surface-emitting lasers (VECSELs)~\cite{a200513.10}, the power losses in the pump process (steps  \ding{172} -- \ding{174}, contributing to 32.4\% power loss) can be removed. Loss in step \ding{175} can be reduced by adjusting the active shape of the gain medium (for example, using VECSEL arrays), which can improve the overlap efficiency. Loss in step~\ding{176} depends on the gain medium material, which generates heat. The pump frequency also affects this efficiency. For instance, $880$~nm for Nd:YVO$_4$ generates less heat than $808$~nm. The diffraction loss in step \ding{177} is very small in our cavity, as the SSLR operates in the stable regime; yet, the radius of the optical devices should be much greater than the TEM$_{00}$ mode radius, which is easy to be satisfied. The transmittance of each optical element  with AR coatings is up to $99\%$. Although it is hard to further improve the optical elements, reducing the intra-cavity beam power (decreasing the output mirror reflectivity and meanwhile improving the gain medium amplification) can also reduce the energy consumption (in steps \ding{177} -- \ding{179}) of the intra-cavity optical components. The energy consumption of the air in step \ding{179} can be neglected in indoor scenario. As for step \ding{180}, the PV efficiency has been continuously improved in recent years. For such single-wavelength applications, the PV efficiency is reported as up to $50\%$ at  $1064$~nm~\cite{a211103.01} or $60\%$ at $875$~nm~\cite{a211103.02}. Above all, many losses can be reduced in future by optimizing the system parameters and also by developing new SSLR structures and materials. The overall PT efficiency in future is expected to reach $15\%$ -- $35\%$.
	
	
\subsection{Misalignment}
\begin{figure}[t]
	\centering
	\includegraphics[width=2.6in]{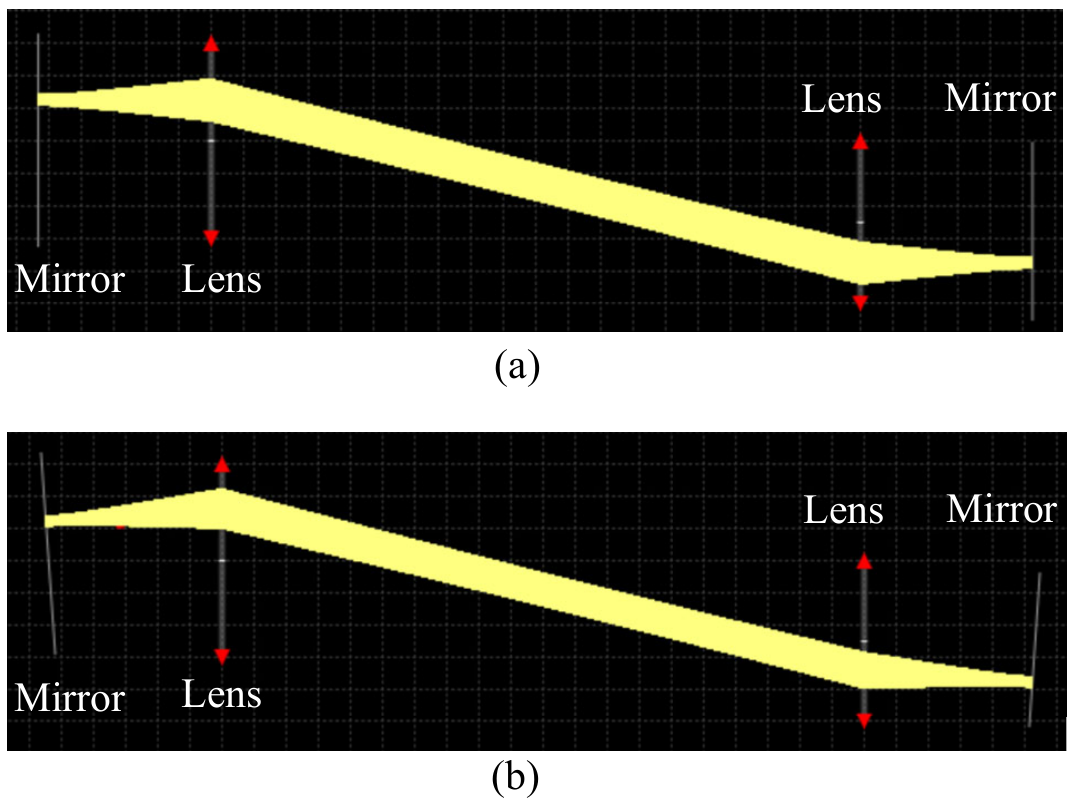}
	\caption{Misalignment of lens and mirror: (a) Lens and mirror inside the retroreflector are set in parallel; (b) mirrors are tilted}
	\label{fig:misalign}
\end{figure}

	Different from a two-mirror resonator which needs to place the mirrors in parallel,  retroreflectors in the SSLR do not need to be aligned as long as they locate in the FOV of the opposite one. The resonant beam is self-aligned between the transmitter and the receiver. As depicted in Fig.~\ref{fig:misalign}(a), even if the two retroreflectors are not coaxially placed, rays can be captured by the resonator, forming an intra-cavity beam (ray optics simulation using open source software~\cite{rayoptic}). The misalignment between the mirror and the lens in a retroreflector is also worth  investigating. As shown in Fig.~\ref{fig:misalign}(b), the mirrors are tilted, while the rays in the cavity can still be captured. A little misalignment between the lens and the mirror in the retroreflectors does not affect the self-producing capability of the intra-cavity beam, although it changes the FOV of the transmitter and the receiver. The tilt angle of the mirror/lens in each TCR should be as small as possible to ensure a large moving range.

\section{Conclusions}
\label{sec:con}
In this paper, we proposed a resonant-beam-based simultaneous information and power transfer~(RB-SWIPT) system which uses spatially separated laser resonator~(SSLR) and second harmonic generation~(SHG)  to provide a high-power and high-capacity mobile transmission channel. We designed an optical frequency-splitting receiver structure, in which the   power  and  information are transferred independently via the resonant beam  and its frequency-doubled beam  while  sharing the same optical transmission path that is self-aligned without positioning and beam steering. We established the intra-cavity power coupling model of the resonant beam and the frequency doubled beam 
as well as the power receiving model to estimate the received battery charging power and the  achievable rate. Numerical results show that the RB-SWIPT system can achieve watt-level battery charging and over $10$-bit/s/Hz achievable rate at $8$-m distance, which can enable permanent battery life and high-quality multimedia stream transmission of mobile devices in indoor scenarios.

There are still several issues to be addressed in future. For instance,
the energy efficiency needs to be improved. Beside, analytical model that considers the incidence angle should be investigated. Furthermore, to improve the performance, a parameter optimization algorithm should be developed.



%

%



\ifCLASSOPTIONcaptionsoff
  \newpage
\fi




\bibliographystyle{IEEETran}
\small
%
\bibliography{mybib}
%
%

%

%
%







\end{document}